\documentclass[10pt,journal]{IEEEtran}

\usepackage{amssymb}
\usepackage{amsmath}
\usepackage{cite}
\usepackage{url}
\usepackage{xcolor}
\usepackage{cite,graphicx,amsmath,amssymb}
\usepackage{fancyhdr}
\usepackage{mdwmath}
\usepackage{mdwtab}
\usepackage{caption}
\usepackage{amsthm}
\usepackage{setspace}
\usepackage{hyperref}
\usepackage{algorithm}
\usepackage{algorithmic}
\usepackage{pdfpages}
\usepackage{mathtools}
\usepackage{subcaption}
\usepackage{makecell}
\usepackage{bm}
\hypersetup{colorlinks=false}

\graphicspath{{./src/img/}}

\newtheorem{remark}{Remark}
\newtheorem{theorem}{Theorem}

\newtheorem{lemma}{Lemma}

\newtheorem{corollary}{Corollary}

\allowdisplaybreaks
\title{TTD Configurations for Near-Field Beamforming: Parallel, Serial, or Hybrid?}
\author{

        Zhaolin~Wang,~\IEEEmembership{Graduate Student Member,~IEEE,}
        Xidong~Mu,~\IEEEmembership{Member,~IEEE,} \\
        Yuanwei~Liu,~\IEEEmembership{Fellow,~IEEE}, and Robert Schober,~\IEEEmembership{Fellow,~IEEE} 
\thanks{Zhaolin Wang, Xidong Mu, and Yuanwei Liu are with the School of Electronic Engineering
and Computer Science, Queen Mary University of London, London E1 4NS,
U.K. (e-mail: zhaolin.wang@qmul.ac.uk, xidong.mu@qmul.ac.uk, yuanwei.liu@qmul.ac.uk).}
\thanks{Robert Schober is with the Institute for Digital Communications, Friedrich-Alexander-Universität Erlangen-Nürnberg, 91054 Erlangen, Germany (e-mail: robert.schober@fau.de).}
\vspace{-0.5cm}
}

\begin{document}

\maketitle
\begin{abstract}
    True-time delayers (TTDs) are popular components for hybrid beamforming architectures to combat the spatial-wideband effect in wideband near-field communications. In this paper, a \emph{serial} and a \emph{hybrid serial-parallel} TTD configuration are investigated for hybrid beamforming architectures. Compared to the conventional parallel configuration, the serial configuration exhibits a \emph{cumulative} time delay caused by multiple TTDs, which potentially alleviates the maximum delay requirements on the individual TTDs. However, independent control of individual TTDs becomes impossible in the serial configuration. Therefore, a hybrid TTD configuration is proposed as a compromise solution. Furthermore, a power equalization approach is proposed to address the cumulative insertion loss of the serial and hybrid TTD configurations. Moreover, the wideband near-field beamforming design for different configurations is studied to maximize the spectral efficiency in both single-user and multiple-user systems. 1) For single-user systems, a closed-form solution for the beamforming design is derived. The \textcolor{black}{preferred user locations} and the required maximum time delay of each TTD configuration are characterized. 2) For multi-user systems, a penalty-based iterative algorithm is developed to obtain a stationary point of the spectral efficiency maximization problem for the considered TTD configurations. In addition, a hybrid-forward-and-backward (HFB) implementation is proposed to enhance the performance of the serial configuration. Our numerical results confirm the effectiveness of the proposed designs and unveil that i) compared to the conventional parallel configuration, both the serial and hybrid configurations can significantly reduce the maximum time delays required for the individual TTDs and ii) the hybrid configuration excels in single-user systems, while the HFB serial configuration is preferred in multi-user systems.
\end{abstract}
\section{Introduction}

The rapid development of wireless applications has led to new requirements for future wireless systems beyond 2030, which are expected to be met by the sixth generation (6G) of wireless networks. The research community has shown great interest in exploring the potential of 6G networks, which aim to offer significant improvements over the current 5G networks \cite{saad2019vision, zhang20196g, dang2020should}. Some of the key features of 6G networks include a peak data rate of terabits per second (Tb/s), which is 1000 times higher than in 5G, a latency of less than one millisecond, which is 10 times lower than in 5G, and an end-to-end reliability of 99.99999\%. In order to meet the exacting demands of 6G wireless networks, the utilization of extremely large-scale multiple-input multiple-output (XL-MIMO) technology, along with the harnessing of the millimeter wave (mmWave) and terahertz (THz) frequency bands, is poised to play a pivotal role \cite{saad2019vision, zhang20196g, dang2020should}. However, the electromagnetic (EM) characteristics of wireless systems should be reevaluated to harness these advanced techniques efficiently. Specifically, the EM radiation field emitted by antennas can be divided into two distinct regions: the far-field and near-field regions \cite{kraus2002antennas}. These regions play a crucial role in wireless communications, determining the behavior of signals as they propagate through space. In the near-field region, which is in the proximity of the antennas, spherical waves are dominant, resulting in complex and often non-uniform radiation patterns. In contrast, the far-field region emerges at a greater distance from the antennas, where the radiation pattern can be characterized by simple planar waves, making it the region of interest for long-range wireless communication in the earlier generations of wireless technologies. However, in 6G wireless networks, the near-field region can span hundreds of meters from the antennas due to the extremely large aperture of antenna arrays and the extremely high carrier frequencies \cite{cui2022near, liu2023near}. Consequently, the design of next-generation wireless technology cannot only rely on the far-field assumption. Instead, the near-field effect becomes an imperative and indispensable factor to be thoroughly considered for 6G.

While the EM radiation field is more complicated in the near-field region compared to the far-field region, it also opens new opportunities for enhancing the performance of wireless communications, including but not limited to near-field beamfocusing \cite{zhang2022beam}, accurate localization \cite{wang2023near}, and augmentation of the degrees of freedom (DoFs) \cite{10262267}.
To harness these advantages of near-field communications, the utilization of extremely large-scale antenna arrays (ELAAs) is crucial to create a sufficiently large near-field region. In this context, a hybrid analog and digital antenna architecture is preferred to attain the required system capacity at minimum hardware and energy cost \cite{heath2016overview}. Conventionally, phase shifters (PSs) are employed to implement the analog part. However, in wideband XL-MIMO systems, the spatial-wideband effect and the frequency-independent nature of PSs lead to the problematic \emph{beam split} effect \cite{liu2023near_tutorial}. \textcolor{black}{For the widely used orthogonal frequency-division multiplexing (OFDM) transmission format,} this effect causes the beams at the majority of the subcarriers to be defocused at the desired location, resulting in unacceptable performance losses, which renders the conventional PS-based hybrid beamforming architecture to be unsuitable for near-field wideband communication. As a remedy, true-time delayers (TTDs) have been proposed as promising analog components to effectively address this challenge \cite{rotman2016true}. \textcolor{black}{In contrast to simple PSs, TTDs use a time delay to realize beamforming.} In particular, TTDs 
allow the alignment of the transmit signals over the entire frequency range of interest, thus mitigating the beam split effect in wideband systems.

\subsection{Prior Works}
Several architectures exploiting TTDs have been proposed for wideband radar and communication systems. \textcolor{black}{One straightforward approach is to simply substitute the PSs in existing hybrid beamforming architectures with TTDs \cite{rotman2016true, longbrake2012true, spoof2020true}.} Although the resulting architecture can achieve flexible wideband beamforming, it leads to unaffordable costs due to the high power consumption and hardware complexity of TTDs, particularly for high carrier frequencies. As a remedy, rather than outright replacing the PSs, the authors of \cite{gao2021wideband} and \cite{dai2022delay} proposed to insert a limited number of TTDs between the PSs and radio-frequency (RF) chains and optimized the corresponding wideband far-field beamsteering design. It was demonstrated that such an architecture is capable of achieving near-optimal performance while keeping the power consumption at a minimum. \textcolor{black}{In pursuit of further advancement, the optimization of the PS and TTD coefficients was explored in \cite{nguyen2022joint} and \cite{ratnam2022joint}, aiming to minimize the matching error with an idealized beamformer. The authors of \cite{dovelos2021channel} employed TTDs to facilitate wideband channel estimation in the presence of the beam split effect, where an efficient estimator based on orthogonal matching pursuit was proposed. To further reduce power consumption, a wideband hybrid beamforming architecture using TTDs with fixed time delays was conceived in \cite{yan2022energy}.}

However, the above works assume far-field planar-wave propagation, which may lead to significant performance losses when ELAAs are employed. As a remedy, the authors of \cite{cui2021near} studied a TTD-based hybrid architecture for near-field beamfocusing, where a piecewise-far-field approximation was proposed for wideband beamforming design. As a further advance, two optimization frameworks, namely fully-digital approximation and a heuristic two-stage design, were developed in \cite{wang2023beamfocusing} to facilitate wideband near-field beamfocusing with TTD-based hybrid architectures. In the aforementioned works, the TTDs are configured in a parallel manner, potentially resulting in an underutilization of their full capability. As a novel contribution, the authors of \cite{zhai2020thzprism} proposed the serial configuration of TTDs, enabling the accumulation of the time delays of multiple TTDs, and conceived a far-field beam spreading approach to extend beam coverage. Most recently, the authors of \cite{najjar2023hybrid} investigated the beamforming design for the serial TTD configuration in wideband far-field communication systems.

\begin{figure*}[t!]
    \centering
    \includegraphics[width=0.7\textwidth]{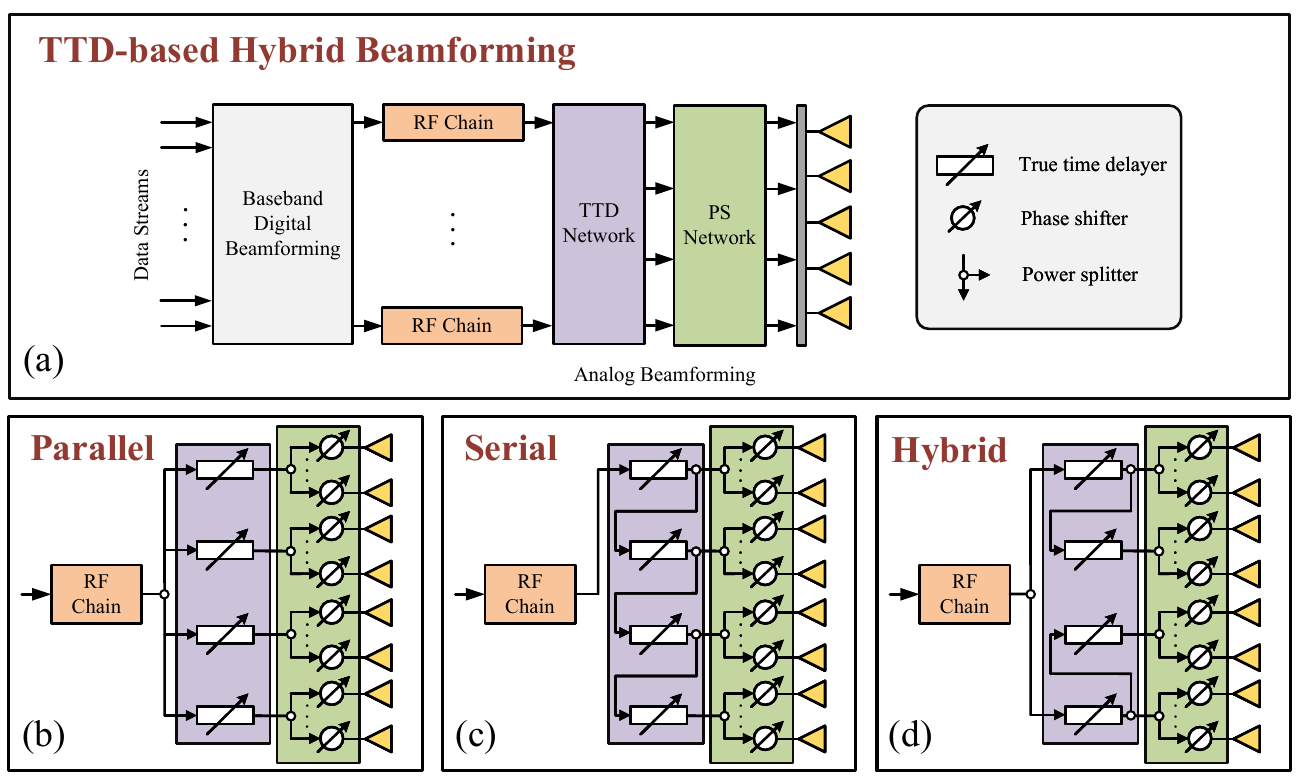}
    \caption{\textcolor{black}{Illustration of the considered \emph{parallel}, \emph{serial}, and \emph{hybrid} configurations of TTD-based hybrid beamforming architectures. In the depicted example, each RF chain is connected to $Q = 4$ TTDs. }}
    \label{fig:model}
\end{figure*}

\subsection{Motivation and Contributions}
The parallel and serial configurations are the two most common configurations for TTD-based hybrid beamforming in the literature, each accompanied by its distinctive merits and drawbacks. \textcolor{black}{In particular, the parallel configuration enables the independent control of individual TTDs, while requiring that each TTD can support large time delays, typically on the order of hundreds or even thousands of picoseconds (ps) \cite{gao2021wideband, dai2022delay, nguyen2022joint, ratnam2022joint, wang2023beamfocusing}. This excessive delay requirement can engender a significant escalation of the costs associated with implementing small-footprint TTDs \cite{yan2022energy, jeong20186, xiang2017low}.} On the contrary, the serial configuration exhibits the potential to remarkably reduce the required time delays of the individual TTDs as it facilitates the accumulation of time delays \cite{zhai2020thzprism, najjar2023hybrid}. However, as the output delay of a specific TTD becomes contingent on the accumulation of all preceding TTDs, the resulting coupling makes it impossible to independently control individual TTDs. Although the performance of the parallel configuration has been investigated for near-field communications \cite{cui2021near, wang2023beamfocusing}, the following question is still unaddressed: \emph{Is the serial configuration more efficacious than the parallel configuration for near-field communications?} Furthermore, to pursue a compromise between the parallel and serial configurations, another interesting question arises: \emph{Could hybrid serial-parallel TTD configurations yield benefits for wideband near-field communications?} Motivated by these two questions, this paper investigates the beamforming design for the parallel, serial, and hybrid configurations and compares their performance in near-field communications. The main contributions of this paper can be summarized as follows:
\begin{itemize}
    \item We investigate serial and hybrid serial-parallel configurations for TTD-based hybrid beamforming in wideband near-field communications. We first analyze the insertion loss of each TTD configuration and propose an equalization method to mitigate the cumulative insertion loss caused by the serial configuration. Then, we investigate the beamforming optimization problem for each configuration for both single- and multi-user systems, with the aim to maximize the spectral efficiency.
    \item \textcolor{black}{We first derive a closed-form beamforming solution for each of the considered configurations in the single-user case. Moreover, we characterize the preferred user locations and the required maximum time delay for each configuration.}
    \item We then propose a penalty-based iterative algorithm to solve the multi-user beamforming optimization problem for each configuration. Specifically, the updates of the optimization variables in each iteration can be obtained either in closed-form or via a simple one-dimensional search. Additionally, based on the results obtained for the single-user case, we propose a tailored hybrid-forward-and-backward (HFB) implementation for enhancing the average performance of the serial TTD configuration in multi-user systems.
    \item We provide numerical results to evaluate the performance of different TTD configurations. It is unveiled that: \textcolor{black}{1) When the maximum achievable time delay of the individual TTDs is limited, the hybrid configuration exhibits the best performance in single-user systems, while the HFB serial configuration performs best in multi-user systems. 2) The parallel configuration requires large TTD time delays to match the performance of the serial and hybrid configurations. However, if such large TTD time delays are affordable, it achieves a higher performance. 3) The proposed equalization method can efficiently mitigate the impact of the cumulative insertion loss of the serial and hybrid TTD configurations.}
\end{itemize}

\subsection{Organization and Notations}
The rest of this paper is organized as follows: Section \ref{sec:system_model} presents the considered parallel, serial, and hybrid configurations and analyzes the corresponding insertion losses. Sections \ref{sec:single} and \ref{sec:multiple} investigate the beamforming design for single-user and multi-user systems, respectively. Section \ref{sec:results} provides numerical results comparing the performance of different configurations. Section \ref{sec:conclusion} concludes this paper.

\emph{Notations:} Scalars, vectors, and matrices are represented by lower-case, bold-face lower-case, and bold-face upper-case letters, respectively. \textcolor{black}{The absolute value of a scalar $x$ is denoted by $|x|$.}  The transpose, conjugate transpose, pseudo-inverse, and trace of a matrix $\mathbf{X}$ are denoted by $\mathbf{X}^T$, $\mathbf{X}^H$, $\mathbf{X}^\dagger$, and $\mathrm{tr}(\mathbf{X})$, respectively. The Euclidean norm of vector $\mathbf{x}$ is denoted as $\|\mathbf{x}\|$, while the Frobenius norm of matrix $\mathbf{X}$ is denoted as $\|\mathbf{X}\|_F$. \textcolor{black}{The $i$-th entry of vector $\mathbf{x}$ is denoted as $[\mathbf{x}]_i$. The entry at the $i$-th row and $j$-th column of matrix $\mathbf{X}$ is denoted as $[\mathbf{X}]_{i,j}$.} A block diagonal matrix with diagonal blocks $\mathbf{x}_1,\dots,\mathbf{x}_N$ is denoted as $\mathrm{blkdiag}\{\mathbf{x}_1,\dots,\mathbf{x}_N\}$. $\mathbb{E}[\cdot]$ represents statistical expectation, while $\mathrm{Re}\{\cdot\}$ refers to the real part of a complex number.
$\mathcal{CN}(\mu, \sigma^2)$ denotes the circularly symmetric complex Gaussian random distribution with mean $\mu$ and variance $\sigma^2$. Symbol $\angle$ represents the phase of a complex value. \textcolor{black}{$\lfloor \cdot \rfloor$ and $\mathrm{exp}(\cdot)$ denote the floor function and the natural exponential function, respectively.} \textcolor{black}{For a vector $\mathbf{x} = [x_1,\dots,x_N]^T$, $\mathrm{exp}(\mathbf{x})$ denotes the vector $[\mathrm{exp}(x_1),\dots,\mathrm{exp}(x_N)]^T$.}  \textcolor{black}{$\mathcal{I}(M)$ refers to the set $\{1,2,\dots,M\}$.}
\section{System Model and Problem Formulation} \label{sec:system_model}

We consider a wideband XL-MIMO communication system. The base station (BS) is equipped with a large-scale uniform linear array (ULA) consisting of $N$ antennas, and uses the mmWave or THz frequency bands to serve $K$ single-antenna users located in the near-field region. \textcolor{black}{Due to the utilization of the large antenna array and the large bandwidth, the considered system inherently suffers from a dual-wideband effect, namely the \emph{frequency-wideband effect} and the \emph{spatial-wideband effect} \cite{gao2021wideband}.} To mitigate the frequency-wideband effect, OFDM with $M$ subcarriers is employed. Let $B$ and $f_c$ denote the system bandwidth and the center frequency, respectively. Accordingly, the $m$-th OFDM subcarrier has a frequency of $f_m = f_c + \frac{B(2m-1-M)}{2M}$. Furthermore, to deal with the spatial-wideband effect, the TTD-based hybrid beamforming architecture illustrated in Fig. \ref{fig:model}(a) is adopted as an energy-efficient solution. Compared to the conventional hybrid beamforming architecture, an additional TTD network is inserted between the PS network and the RF chains. According to the property of the Fourier transform, a time delay $t$ introduced by a TTD is transformed into a phase shift $e^{-j 2 \pi f_m t}$ on the $m$-th subcarrier, thus facilitating \emph{frequency-dependent} analog beamforming.

\subsection{Signal Model}
We assume that the TTD-based hybrid beamforming architecture employs $N_{\text{RF}}$ RF chains. The output signal of each RF chain is fed to all antenna elements through $Q$ TTDs and $N$ PSs. Furthermore, the set of $N$ PSs is partitioned into $Q$ distinct groups, each consisting of $N_{\text{sub}} = N/Q$ PSs connected to one TTD. \textcolor{black}{For the $n$-th RF chain, let $\tilde{t}_{n,q}$ denote the time delay of the output signal of the $q$-th TTD and let $\mathbf{a}_{n,q} \in \mathbb{C}^{N_{\text{sub}} \times 1}$ denote the vector containing the coefficients of the PSs connected to the $q$-th TTD.} As PSs can only adjust the phase, the following constraint \textcolor{black}{has to be satisfied:}
\begin{equation} \label{constraint_PS}
    \left| \left[\mathbf{a}_{n,q}\right]_i \right| = 1, \forall i \in \mathcal{I}(N_{\text{sub}}).
\end{equation}
On the $m$-th subcarrier, the frequency-domain phase shift realized by time delay $\tilde{t}_{n,q}$ is given by $e^{-j 2 \pi f_m \tilde{t}_{n,q}}$. Here, we assume that all antennas are fed with the same power by adjusting the coefficients of the power splitters depicted in Fig. \ref{fig:model}, which will be detailed in the following sections. As a result, the matrix representing the overall analog beamformer $\bar{\mathbf{A}}_m \in \mathbb{C}^{N \times N_{\text{RF}}}$ on the $m$-th subcarrier for the $N_{\text{RF}}$ RF chains can be expressed as 
\begin{align}
    &\bar{\mathbf{A}}_m = \frac{1}{\sqrt{N}} \times \nonumber\\
    &\begin{bmatrix}
        \mathbf{a}_{1,1} \mathrm{exp}\left(-j 2\pi f_m \tilde{t}_{1,1}\right) \!\!\!\! & \dots & \!\!\!\!\mathbf{a}_{N_{\text{RF}},1} \mathrm{exp}\left(-j 2\pi f_m \tilde{t}_{N_{\text{RF}},1}\right) \\
        \vdots \!\!\!\!& \ddots &\!\!\!\! \vdots \\
        \mathbf{a}_{1,Q} \mathrm{exp}\left(-j 2\pi f_m \tilde{t}_{1,Q}\right) \!\!\!\!& \dots &\!\!\!\!\mathbf{a}_{N_{\text{RF}},Q} \mathrm{exp}\left(-j 2\pi f_m \tilde{t}_{N_{\text{RF}},Q}\right)
    \end{bmatrix}.
\end{align}  
Matrix $\bar{\mathbf{A}}_m$ can be compactly written as
\begin{equation}
    \bar{\mathbf{A}}_m = \mathbf{A} \mathbf{T}_m,
\end{equation} 
where $\mathbf{A} \in \mathbb{C}^{N \times Q N_{\text{RF}}}$ and $\mathbf{T}_m \in \mathbb{C}^{Q N_{\text{RF}} \times N_{\text{RF}}}$ represent the PS-based analog beamformer and the TTD-based analog beamformer, respectively. Specifically, matrix $\mathbf{A}$ is given by 
\begin{equation}
    \mathbf{A} = \left[ \tilde{\mathbf{A}}_1, \dots, \tilde{\mathbf{A}}_{N_{\text{RF}}} \right],
\end{equation}
where $\tilde{\mathbf{A}}_n = \frac{1}{\sqrt{N}} \mathrm{blkdiag}\left\{\mathbf{a}_{n,1},\dots,\mathbf{a}_{n, Q}\right\}$. 
Matrix $\mathbf{T}_m$ is given by
\begin{align} \label{eqn_Tm}
    \mathbf{T}_m = &\mathrm{blkdiag}\left\{\mathrm{exp}\left(-j2\pi f_m \mathbf{t}_1\right), \dots, \mathrm{exp}\left(j2\pi f_m \mathbf{t}_{N_{\text{RF}}}\right) \right\} \nonumber \\
    \triangleq & F(f_m, \mathbf{t}_1,\dots,\mathbf{t}_{N_{\text{RF}}}),
\end{align} 
where $\mathbf{t}_n = [\tilde{t}_{n,1},\dots,\tilde{t}_{n,Q}]^T$ is the time delay vector.
By exploiting the TTD-based hybrid beamforming architecture, the transmit signal on the $m$-th subcarrier is given by
\begin{equation}
    \mathbf{x}_m = \mathbf{A} \mathbf{T}_m \mathbf{D}_m \mathbf{s}_m = \mathbf{A} \mathbf{T}_m \sum_{k =1}^K \mathbf{d}_{m,k} s_{m,k},
\end{equation}
where $\mathbf{D}_m = [\mathbf{d}_{m,1},\dots,\mathbf{d}_{m,K}] \in \mathbb{C}^{N_{\text{RF}} \times K}$ and $\mathbf{s}_m = [s_{m,1},\dots,s_{m,K}]^T$ denote the baseband digital beamformer and the unit-power information symbols for the $K$ users, respectively. \textcolor{black}{Here, symbol $s_{m,k}$ is intended for user $k$.} We assume that the information symbols for each user are independent and identically distributed (i.i.d.), i.e., $\mathbb{E}[ \mathbf{s}_m \mathbf{s}_m^H ] = \mathbf{I}_K$. Consequently, the received signal of user $k$ on the $m$-th subcarrier can be expressed as 
\begin{align} \label{receive_signal}
    y_{m,k} = &\underbrace{\mathbf{h}_{m,k}^H \mathbf{A} \mathbf{T}_m \mathbf{d}_{m,k} s_{m,k}}_{\text{desired signal}} \nonumber \\ &+ \underbrace{\sum_{i =1, i \neq k}^K \mathbf{h}_{m,k}^H \mathbf{A} \mathbf{T}_m \mathbf{d}_{m,i} s_{m,i}}_{\text{inter-user interference}} + n_{m,k},
\end{align}
where $n_{m,k} \sim \mathcal{CN}(0, \sigma^2)$ and $\mathbf{h}_{m,k} \in \mathbb{C}^{N \times 1}$ denote i.i.d. complex Gaussian noise and the wideband near-field channel of user $k$ on the $m$-th subcarrier, respectively. In this paper, the uniform spherical wave near-field channel model is adopted\footnote{\textcolor{black}{We note that the conventional planar-wave-based far-field channel model is an approximation of the spherical-wave-based near-field channel model \cite{liu2023near_tutorial}. Therefore, the designs proposed in this paper can be easily extended to far-field scenarios.}}, which comprises a line-of-sight (LoS) component and several non-line-of-sight (NLoS) components caused by scatterers. For this model, channel $\mathbf{h}_{m,k}$ is characterized as follows \cite{liu2023near_tutorial}
\vspace{-0.2cm}
\begin{equation}
    \mathbf{h}_{m,k} = \beta_{m,k} \boldsymbol{\alpha}^*(f_m, r_k, \theta_k) + \sum_{l = 1}^{L_k} \tilde{\beta}_{m, k,l} \boldsymbol{\alpha}^*(f_m, \tilde{r}_{k,l}, \tilde{\theta}_{k,l}).
    \vspace{-0.1cm}
\end{equation} 
Here, $L_k$ denotes the number of scatterers affecting the channel between the BS and user $k$. $r_k$ and $\theta_k$ denote the distance and direction of user $k$ with respect to the center of the ULA at the BS, respectively. $\tilde{r}_{k,l}$ and $\tilde{\theta}_{k,l}$ denote the distance and direction of the $l$-th scatterer, respectively, $\beta_{m,k}$ and $\tilde{\beta}_{m,k,l}$ denote the complex channel gains of the LoS and NLoS components, respectively. Vector $\boldsymbol{\alpha}(f, r, \theta) \in \mathbb{C}^{N \times 1}$ is the near-field array response vector, which for a ULA operating at frequency $f$ can be modeled as \cite{liu2023near_tutorial, wang2018spatial}
\begin{equation}
    \boldsymbol{\alpha}(f, r, \theta) = \mathrm{exp}\left(-j \frac{2\pi f}{c} \mathbf{r}(r,\theta)\right).
\end{equation}  
Specifically, $c$ denotes the speed of light and $\mathbf{r}(r,\theta) \in \mathbb{R}^{N \times 1}$ denotes the vector containing the propagation distances between the BS antennas and the user. For spherical wave propagation in the near-field region, the $n$-th element of vector $\mathbf{r}(r,\theta)$ is given by\cite{liu2023near_tutorial}
\begin{equation}
    [\mathbf{r}(r,\theta)]_n = \sqrt{ r^2 + \delta_n^2 d^2 - 2 r \delta_n d \cos \theta },
\end{equation}
where $d = \lambda_c/2 = c/(2f_c)$ denotes the antenna spacing and $\delta_n \triangleq n-1-\frac{N-1}{2}$. \textcolor{black}{In this paper, we focus on theoretical analysis and determining a performance upper-bound of the considered TTD configurations. Therefore, the channel state information is assumed to be perfectly known at the BS.}

Based on \eqref{receive_signal}, the achievable rate of user $k$ on the $m$-th subcarrier can be expressed as
\begin{equation}
    R_{m,k} = \log_2 \left( 1 + \frac{|\mathbf{h}_{m,k}^H \mathbf{A} \mathbf{T}_m \mathbf{d}_{m,k}|^2}{\sum_{i =1, i \neq k}^K |\mathbf{h}_{m,k}^H \mathbf{A} \mathbf{T}_m \mathbf{d}_{m,i}|^2 + \sigma^2 }  \right).
\end{equation}
Thus, the spectral efficiency of the system is given by
\begin{equation}
    R = \frac{1}{M+L_{\text{cp}}} \sum_{m=1}^M \sum_{k=1}^K R_{m,k},
\end{equation}
where $L_{\text{cp}}$ denotes the length of the OFDM cyclic prefix.

\subsection{TTD Configurations}

Conventionally, the TTDs of each RF chain are configured in parallel \cite{gao2021wideband, dai2022delay}, as illustrated in Fig. \ref{fig:model}(b). For this configuration, each TTD independently delays the output signal of the corresponding RF chain, enabling independent control of the individual TTDs. However, a significant performance loss may occur when the maximum delay of the individual TTDs is limited \cite{wang2023beamfocusing}.
To overcome this limitation, a serial configuration of TTDs can be used, as depicted in Fig. \ref{fig:model}(c). For this configuration, the output signal of each RF chain is sequentially fed into the TTDs. As a result, the output signal of each TTD contains the accumulation of the time delays introduced by the preceding TTDs, effectively increasing the range of achievable time delays. However, it is important to note that the serial configuration of TTDs may suffer from certain disadvantages due to the coupling of the delays, which will be discussed in the following sections. Therefore, to potentially harness the benefits of both the parallel and serial configurations, we propose a hybrid configuration, as shown in Fig. \ref{fig:model}(d)\footnote{\textcolor{black}{In this paper, for simplicity of presentation, we focus on a single potential implementation of the hybrid configuration containing two equal-sized parallel groups of serially connected TTDs. However, other possible implementations (e.g., parallel groups of non-equal size) could potentially yield enhanced performance for the hybrid configuration which we will investigate in our future work.}}. In particular, the TTDs are equally divided into two groups. The TTDs within each group are interconnected in a serial manner, while the groups themselves are parallel. Compared to the purely serial configuration, the hybrid configuration reduces the coupling of the TTDs while retaining the capability to realize larger signal delays. Let $t_{n,q}$ denote the time delay introduced by the $q$-th TTD connected to the $n$-th RF chain. Due to the limited range of delays supported by individual TTDs, $t_{n,q}$ has to satisfy the fixed interval constraint \cite{yan2022energy}:
\begin{equation} \label{maximum_delay_constraint}
    t_{n, q} \in [0, t_{\max}], \forall n \in \mathcal{I}(N_{\text{RF}}), q \in \mathcal{I}(Q),
\end{equation}
where $t_{\max}$ denotes the maximum achievable time delay of a TTD. In the following, we provide the exact expressions for $\mathbf{t}_n$ in \eqref{eqn_Tm} and the coefficients of the power splitters for different TTD configurations.

\subsubsection{Parallel Configuration}
As illustrated in Fig. \ref{fig:model}(b), for this configuration, the time delays of the output signals of all TTDs are independent from each other. Therefore, the $q$-th entry of time delay vector $\mathbf{t}_n$ is given by
\begin{equation}
    \label{parallel}
    \tilde{t}_{n,q} = t_{n,q}.
\end{equation} 
Moreover, the power fed to each antenna is determined by the power splitters between the RF chains and the TTDs and between the TTDs and the PSs. To guarantee that each antenna is fed with the same power, the power can be split equally for both cases.

\subsubsection{Serial Configuration}
\begin{figure}[t!]
    \centering
    \includegraphics[width=0.3\textwidth]{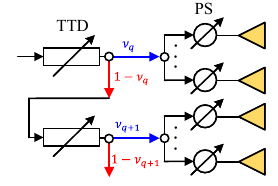}
    \caption{Illustration of the power splitter coefficients for the serial configuration.}
    \label{fig:splitter}
\end{figure}

As shown in Fig. \ref{fig:model}(c), for the serial configuration, the time delay of each TTD's output signal becomes dependent on the delay of the previous TTDs, resulting in a cascade effect. The $q$-th entry of time delay vector $\mathbf{t}_n$ is determined by whether the TTDs are connected in a forward manner (i.e., from the first to the last TTD) or in a backward manner (i.e., from the last to the first TTD). Specifically, for the forward serial configuration, the $q$-th entry of time delay vector $\mathbf{t}_n$ is given by
\begin{equation}
    \label{serial-I_1}
    \tilde{t}_{n,q} = \sum\nolimits_{i=1}^q t_{n,i},
\end{equation}
while for the backward serial configuration, it is given by
\begin{equation}
    \label{serial-I_2}
    \tilde{t}_{n,q} = \sum\nolimits_{i=q}^{Q} t_{n,i}.
\end{equation}
Furthermore, in the serial configuration, the power fed to each antenna is also subject to a cascade effect of the power splitters located at the output of the TTDs, as shown in Fig. \ref{fig:splitter}. In this case, splitting the power equally is no longer suitable. Let us take the forward serial configuration as an example. If the power is split equally, the output power of the first TTD is $1/2$ of the input power from the RF chain, while that of the last TTD diminishes to only $1/2^Q$ of the input power, which renders the antennas connected to the last TTD almost unusable for transmission. As a remedy, an unequal power splitter \cite{du2016unequal} should be used, which is capable of dividing the input power into desired fractions by adjusting its impedance. Let $\nu_q$ denote the power splitting coefficient, i.e., the fraction of power allocated to the $q$-th TTD by the $q$-th power splitter, as depicted in Fig. \ref{fig:splitter}. Then, the effective output power of the $q$-th TTD can be expressed as 
\begin{subequations} \label{output_power}
    \begin{align}
        P_{\text{out},1} = &\nu_1 P_{\text{in}},\\
        P_{\text{out},q} = &\nu_q \prod_{i=1}^{q-1} (1-\nu_i) P_{\text{in}}, \forall q = 2,\dots,Q,
    \end{align} 
\end{subequations}
where $P_{\text{in}}$ denotes the input power of the first TTD.
Then, to guarantee equal output power for all TTDs, i.e., $P_{\text{out},1} = P_{\text{out},2} = \dots = P_{\text{out},Q} = P_{\text{in}}/Q$, $\nu_q$ has to satisfy
\begin{equation} \label{solution_non_equ}
    \nu_q = \frac{1}{Q-q+1}.
\end{equation}
The value of $\nu_q$ for the backward serial configuration can be designed similarly, which thus is omitted here.

\subsubsection{Hybrid Configuration}
For the hybrid configuration, the TTDs are equally divided into two groups, as shown in Fig. \ref{fig:model}(d). In the first group, the TTDs are connected in a forward serial manner, whereas in the second group, the TTDs are connected in a backward serial manner. Such a configuration has the potential to benefit the average performance, as will be shown in Sections \ref{sec:single} and \ref{sec:results}. Therefore, by assuming $Q$ is an even number, the $q$-th entry of the corresponding time delay vector $\mathbf{t}_n$ is given by 
\begin{equation}
    \label{serial-II}
    \tilde{t}_{n,q} = \begin{cases}
        \sum_{i=1}^{q} t_{n,i}, &q \le \frac{Q}{2}, \\
        \sum_{i=q}^{Q} t_{n,i}, &q > \frac{Q}{2}.
    \end{cases}
\end{equation}
Additionally, the output power of each RF chain is first divided equally between the two parallel groups of TTDs by the power splitter between the RF chain and the TTDs. Subsequently, within each group of TTDs, the power splitting coefficient is designed in the same way as for the serial configuration.

\begin{remark} \label{remark_1}
    \emph{For the serial configuration, the time delays in time delay vector $\mathbf{t}_n$ are either monotonically increasing or decreasing with respect to TTD index $q$. Conversely, for the hybrid configuration, the time delays initially increase and subsequently decrease. These distinct characteristics of the serial and hybrid configurations make them suitable for users located at different positions, which will be discussed in the following sections.}
\end{remark}

\begin{remark} \label{remark_2}
    \emph{For infinite-range TTDs, i.e., when $t_{\max} = +\infty$, both the serial and hybrid configurations can be considered as special cases of the parallel configuration. This is because, in this scenario, any time delay vector realized by the serial and hybrid configurations can also be achieved by the parallel configuration. However, the reverse is not true due to the monotonicity constraints discussed in \textbf{Remark \ref{remark_1}}. On the contrary, for practical finite-range TTDs, i.e., when $t_{\max}$ is finite, the serial and hybrid configurations are no longer special cases of the parallel configuration. For instance, the maximum time delay achieved by the serial configuration is $Q \times t_{\max}$, while that of the parallel configuration is limited to $t_{\max}$.}
\end{remark}

\subsection{Insertion Loss and Power Equalization}
In contrast to baseband digital signal processing, analog signal processing introduces additional insertion losses as the signal traverses through the analog devices \cite{gao2021wideband}. For the scope of this paper, our focus lies on quantifying the insertion loss caused specifically by the TTDs and the power splitters, while neglecting the losses caused by other elements, as they are identical for all considered TTD configurations. While the serial and hybrid configurations offer the potential to achieve longer time delays, they also incur a cumulative insertion loss. Specifically, denoting the insertion losses of the TTDs and the power splitters by $\eta_1 \ge 1$ and $\eta_2 \ge 1$, respectively, for the parallel configuration, the insertion loss manifested in the output signal of the $q$-th TTD is given by $\eta_{\text{parallel}}(q) = \eta$, where $\eta \triangleq \eta_1 \eta_2$. In contrast, for the forward serial configuration, it is given by $\eta_{\text{serial}}(q) = \eta^q$, which for $\eta \ge 1$ is greater compared to the parallel configuration. As a result, the antenna sub-array connected to the TTDs at the end of the serial configuration may not operate effectively due to the lower output power caused by the cumulative insertion loss. To address this issue, the coefficient of the power splitter $\nu_q$ should be redesigned to equalize the output power of each TTD taking into account the effect of the cumulative insertion loss. 

Let us take the forward serial configuration as an example. Taking into account the insertion loss, the output power in \eqref{output_power} becomes
\begin{subequations}
    \begin{align}
        P_{\text{out},1} = &\frac{1}{\eta} \nu_1 P_{\text{in}}, \\
        P_{\text{out},q} = &\frac{1}{\eta^q} \nu_q \prod_{i=1}^{q-1} (1-\nu_i) P_{\text{in}}, \forall q = 2,\dots,Q.
    \end{align}
\end{subequations}
To obtain the value of $\nu_q$ needed such that $P_{\text{out},1} = P_{\text{out},2} = \dots = P_{\text{out},Q} \triangleq \bar{P}_{\text{out}}$, we define auxiliary variables $\tilde{\nu}_1 \triangleq \nu_1$ and $\tilde{\nu}_q \triangleq \nu_q \prod_{i=1}^{q-1} (1-\nu_i)$, which satisfy the following two conditions:
\begin{equation}
    \sum_{q=1}^Q \tilde{\nu}_q = 1, \quad \frac{\tilde{\nu}_q}{\eta^q} P_{\text{in}} = \bar{P}_{\text{out}}.
\end{equation}
Based on these conditions, it can be readily shown that 
\begin{equation}
    \tilde{\nu}_q = \frac{\eta^{q-1} - \eta^q}{1 - \eta^Q}.
\end{equation}
Then, according to the definition of $\tilde{\nu}_q$, the value of the coefficient of the power splitter that equalizes the output power of the $q$-th TTD is given by 
\begin{equation} \label{solution_eqn}
    \nu_q = \frac{1 - \eta}{1 - \eta^{Q-q+1}}.
\end{equation}
Furthermore, the effective insertion loss of each TTD in the serial configuration after equalization can be calculated as 
\begin{equation} \label{eqn:insertion_serial}
    \tilde{\eta}_{\text{serial}} = \frac{\frac{1}{Q} P_{\text{in}}}{\bar{P}_{\text{out}}} = \frac{\eta^q}{\tilde{\nu}_q Q} = \frac{1 - \eta^Q}{(1 - \eta) Q} \times \eta.
\end{equation}

The coefficients of the power splitter in the backward serial configuration and the hybrid configuration can be calculated similarly. In particular, the effective insertion loss in the backward serial configuration after equalization is the same as \eqref{eqn:insertion_serial}, while that in the  hybrid configuration is given by 
\begin{equation}
    \tilde{\eta}_{\text{hybrid}} = \frac{1 - \eta^{\frac{Q}{2}} }{(1 - \eta) \frac{Q}{2}} \times \eta.
\end{equation}
\textcolor{black}{We note that we neglect the insertion loss caused by the power splitter between the RF chain and the TTDs as its impact is marginal compared to the considered cumulative losses.}

It is also worth noting that even after equalization, the insertion losses of the serial and hybrid configurations are still enlarged by factors of $\frac{1 - \eta^Q}{(1 - \eta) Q}$ and $\frac{1 - \eta^{\frac{Q}{2}} }{(1 - \eta) \frac{Q}{2}}$, respectively, compared to that of the parallel configuration. Therefore, the serial and hybrid configurations have more stringent requirements regarding the insertion loss. Furthermore, the insertion loss can be readily compensated for by introducing additional power amplifiers. Since the insertion loss has no direct impact on the beamforming design, in the subsequent sections, we neglect the insertion loss for beamforming design. Nevertheless, we do take into account its impact for the numerical results in Section \ref{sec:results}.

\subsection{Problem Formulation}
\textcolor{black}{To compare the performance of the considered parallel, serial, and hybrid TTD configurations, we focus on the maximization of the spectral efficiency with respect to the analog beamformers $\mathbf{A}$ and $\mathbf{T}_m$ and the baseband digital beamformer $\mathbf{D}_m$. The resulting optimization problem is given by 
\begin{subequations} \label{problem:SE_max}
    \begin{align}
        \max_{\mathbf{A}, \mathbf{T}_m, \mathbf{D}_m} \quad & \sum_{m=1}^M \sum_{k=1}^K R_{m,k}, \\
        \label{constraint:power}
        \mathrm{s.t.} \quad & \|\mathbf{A} \mathbf{T}_m \mathbf{D}_m\|_F^2 \le P_t, \forall m \in \mathcal{I}(M), \\
        \label{constraint:analog}
        & \left|[\mathbf{A}]_{i,j} \right| = 1, \forall (i,j) \in \mathcal{F}, \\
        \label{constraint:TTD_1}
        & \mathbf{T}_m = F(f_m, \mathbf{t}_1,\dots,\mathbf{t}_{N_{\text{RF}}}), \forall m \in \mathcal{I}(M), \\
        \label{constraint:TTD_2}
        & \mathbf{t}_n \in \mathcal{T}_n, \forall n \in \mathcal{I}(N_{\text{RF}}), \\
        \label{constraint:TTD_3}
        & t_{n, q} \in [0, t_{\max}], \forall n \in \mathcal{I}(N_{\text{RF}}), q \in \mathcal{I}(Q),
    \end{align}
\end{subequations}
Here, constraint \eqref{constraint:power} limits the transmit power per-subcarrier \cite{sohrabi2017hybrid}. Constraint \eqref{constraint:analog} accounts for the unit-modulus constraint of the PS-based analog beamformer, where $\mathcal{F}$ is the set of non-zero elements of $\mathbf{A}$. Constraints \eqref{constraint:TTD_1}-\eqref{constraint:TTD_3} represent the hardware constraints of the TTD-based analog beamformers. More specifically, $\mathcal{T}_n$ denotes the feasible set of the time delay vector for the TTDs connected to the $n$-th RF chain, which is defined based on the configuration constraints in \eqref{parallel}, \eqref{serial-I_1}, \eqref{serial-I_2}, or \eqref{serial-II} and the TTD configuration employed for this RF chain. In the sequel, we first investigate problem \eqref{problem:SE_max} for signal-user systems to unveil the origin of the performance gains achieved with the serial and hybrid configurations over the parallel configuration and to obtain insights for system design. Then, multi-user systems are studied in Section \ref{sec:multiple}.}

\section{Single-User System} \label{sec:single}
In this section, the single-user system, where $K = 1$, is investigated to draw useful insights. In this case, a single RF chain is equipped at the BS to serve the user, i.e., $N_{\text{RF}} = 1$. Thus, by dropping the user index and the RF chain index, the achievable rate of the user on the $m$-th subcarrier can be simplified as follows
\begin{align}
    R_m = &\log_2 \left( 1 + \frac{1}{\sigma^2} \big|\mathbf{h}_m^H \tilde{\mathbf{A}} \mathrm{exp}\left(-j 2\pi f_m \mathbf{t}\right) d_m \big|^2 \right) \nonumber \\
    \approx & \log_2 \left(1 + \frac{|\beta_m|^2}{\sigma^2} \big|\boldsymbol{\alpha}_m^T \tilde{\mathbf{A}} \mathrm{exp}\left(-j 2\pi f_m \mathbf{t}\right) d_m \big|^2  \right),
\end{align}  
where $d_m$ denotes the digital beamforming coefficient for the $m$-th subcarrier and $\boldsymbol{\alpha}_m \triangleq \boldsymbol{\alpha}(f_m, r, \theta)$. The approximation in the last step stems from the fact that the communication channel is dominated by the LoS path due to the severe scattering loss of the NLoS paths in the mmWave and THz bands. Furthermore, the maximum power constraint in \eqref{constraint:power} can be rewritten as follows
\begin{equation}
    \big\| \tilde{\mathbf{A}} e^{-j 2\pi f_m \mathbf{t}} d_m  \big\|^2 = N |d_m|^2 \le P_t.
\end{equation}
It is easy to prove that $R_m$ is maximized when $N |d_m|^2 = P_t$, and is given by 
\begin{equation} \label{rate_single_user}
    R_m = \log_2 \left(1 + \frac{|\beta_m|^2 P_t}{\sigma^2 N} \big|\boldsymbol{\alpha}_m^T \tilde{\mathbf{A}} \mathrm{exp}\left(-j 2\pi f_m \mathbf{t}\right)\big|^2  \right).
\end{equation}  
\textcolor{black}{In the following, we begin by addressing the spectral efficiency maximization problem for infinite-range TTDs, which establishes a performance upper-bound. Subsequently, leveraging the structure of the optimal beamformers for infinite-range TTDs, we design the beamformers for finite-range TTDs.}   

\subsection{Infinite-Range TTDs}
According to the discussion in \textbf{Remark \ref{remark_2}}, for infinite-range TTDs, the serial and hybrid configurations can be regarded as special cases of the parallel configuration. Therefore, we focus on the parallel configuration in this subsection. Define the time delay vector of the infinite-range TTDs as $\mathbf{t}_{\infty} = [ t_1^{\infty},\dots,t_{Q}^{\infty}]^T$, where $t^{\infty}_q \in [0, +\infty)$. For simplicity, let $2 \pi \psi_{q, i}$ denote the phase introduced by the $i$-th PS connected to the $q$-th TTD and define $\tilde{r}_{q,i} = r_{(q-1)N_{\text{sub}} + i}$. Then, the array gain realized by the PSs and infinite-range TTDs on the $m$-th subcarrier can be expressed as 
\begin{align} \label{single_array_gain}
    &G_m(f, r, \theta) = \left| \boldsymbol{\alpha}^T_m \tilde{\mathbf{A}} \mathrm{exp}\left(-j 2\pi f_m \mathbf{t}_{\infty}\right) \right| \nonumber \\ &= \bigg|\sum_{q=1}^{Q} \sum_{i=1}^{N_{\text{sub}}} \underbrace{\mathrm{exp}\left(-j 2\pi f_m \tilde{r}_{q,i}/c\right)}_{\text{array response}} \underbrace{\mathrm{exp}\left(j 2 \pi (\psi_{q,i} - f_m t^{\infty}_q )\right)}_{\text{analog beamformer}} \bigg|.
\end{align}
Considering \eqref{rate_single_user}, maximizing the spectral efficiency is equivalent to maximizing the array gain $G(f_m, r, \theta)$ for all subcarriers. However, as can be observed, even if infinite-range TTDs are employed, finding the optimal $\psi_{q,i}$ and $t^{\infty}_q$ such that $G(f_m, r, \theta)$ is maximized on all subcarriers is still intractable. As a remedy, we exploit the piecewise-near-field model proposed in \cite{wang2023beamfocusing}. This model exploits the fact that the entire antenna array is divided into $Q$ non-overlapping sub-arrays, each connected to an individual TTD.  Then, the distance between the user and the center of the sub-array connected to the $q$-th TTD is given by  
\begin{equation}
    r_q^{\text{sub}} = \sqrt{ r^2 + \chi_q^2 d^2 - 2 r \chi_q d \cos \theta  },
\end{equation}
where $\chi_q \triangleq (q-1 - \frac{Q-1}{2}) N_{\text{sub}}$. Then, the phase of the array response in \eqref{single_array_gain} can be reformulated as follows:
\begin{align} \label{approx_1}
    &f_m \tilde{r}_{q,i}/c = f_m (\tilde{r}_{q,i}-r_q^{\text{sub}} + r_q^{\text{sub}})/c \nonumber \\
    &= (f_m - f_c)(\tilde{r}_{q,i}-r_q^{\text{sub}})/c + f_c (\tilde{r}_{q,i}-r_q^{\text{sub}})/c +  f_m r_q^{\text{sub}}/c.
\end{align}
\textcolor{black}{Assuming the number of TTDs is sufficiently large, i.e., the size of the sub-arrays is sufficiently small, $\tilde{r}_{q,i} - r_q^{\text{sub}} \ll r_q^{\text{sub}}$ holds.} \textcolor{black}{This fact, coupled with the observation that the absolute difference between $f_m$ and $f_c$ is typically much smaller than their individual values, indicates that the phase of the array response is dominated by the last two terms in \eqref{approx_1}. As a result, the array response can be approximated by
\begin{align} \label{approx_array_response}
    \mathrm{exp}&\left(-j 2\pi f_m \tilde{r}_{q,i}/c\right) \approx \mathrm{exp}\left(-j 2 \pi \vartheta_{m,q,i} \right),
\end{align}
where 
\begin{equation}
    \vartheta_{m,q,i} = f_c (\tilde{r}_{q,i}-r_q^{\text{sub}})/c +  f_m r_q^{\text{sub}}/c.
\end{equation}
The above approximation can be explained from another perspective. In particular, the delay difference $(\tilde{r}_{q,i}-r_q^{\text{sub}})/c$ reflects the beam split effect within a sub-array. Due to the small size of the sub-arrays, the beam split effect within each sub-array is negligible. Based on the above approximation, the phase for each sub-array can be assumed to be frequency-independent.} \textcolor{black}{A detailed analysis of the required number of TTDs and sub-array sizes for \eqref{approx_array_response} to be an accurate approximation can be found in \cite{cui2021near}.} 

Substituting the approximation in \eqref{approx_array_response} into \eqref{single_array_gain} yields
\begin{align}
    G_m \approx &\left|\sum_{q=1}^{Q} \sum_{i=1}^{N_{\text{sub}}} \mathrm{exp}\left(-j 2 \pi \left(  \vartheta_{m,q,i} - \psi_{q,i} + f_m t^{\infty}_q\right) \right) \right| \nonumber \\
    \overset{(a)}{\le} &\sum_{q=1}^{Q} \sum_{i=1}^{N_{\text{sub}}} \left| \mathrm{exp}\left(-j 2 \pi \left(  \vartheta_{m,q,i} - \psi_{q,i} + f_m t^{\infty}_q\right) \right) \right| = N,
\end{align}
where $(a)$ stems from the triangle inequality. It is well known that the condition for $(a)$ to hold with equality is that all terms in the sum have the same phase. Without loss of generality, we set this common phase to $-2\pi f_m \mu$, where $\mu$ is an arbitrary scalar. Then, the following condition has to be satisfied such that the equality holds:
\begin{align}
    \vartheta_{m,q,i} - \psi_{q,i} + f_m t^{\infty}_q = f_m \mu.
\end{align} 
\textcolor{black}{One solution for PS coefficient $\psi_{q,i}$ and TTD coefficient $t_q^{\infty}$ that satisfies the above condition is}
\begin{equation}
    \textcolor{black}{\psi_{q,i} = - f_c (\tilde{r}_{q,i}-r_q^{\text{sub}})/c, \quad t^{\infty}_q = -r_q^{\text{sub}}/c + \mu.}
\end{equation}
The above solution is independent of the subcarrier index $m$, which implies that it approximately maximizes the array gain for all subcarriers \textcolor{black}{when the size of sub-arrays is sufficiently small}. The value of $\mu$ can be adjusted such that $t^{\infty}_q \ge 0, \forall q$, while keeping the maximum value of $t^{\infty}_q$ as small as possible. To this end, $\mu$ can be set to 
\begin{equation}
    \mu = r_{\max}^{\text{sub}}/c,
\end{equation} 
where $r_{\max}^{\text{sub}} = \max \{ r_q^{\text{sub}}, \forall q\}$.
Then, the time delay of the infinite-range TTDs can be designed as follows:
\begin{equation}
    t^{\infty}_q = (r_{\max}^{\text{sub}} - r_q^{\text{sub}})/c.
\end{equation}

\begin{figure}[t!]
    \centering
    \includegraphics[width=0.35\textwidth]{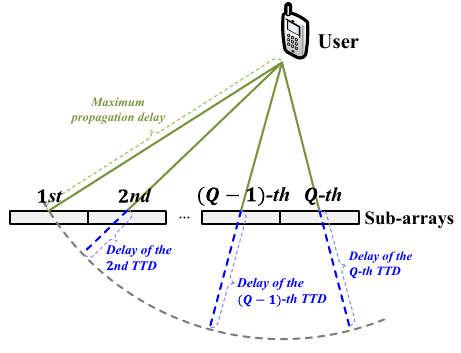}
    \caption{Illustration of the time delay introduced by infinite-range TTDs for mitigating the beam split effect.}
    \label{fig:delay}
\end{figure}

\noindent The above solution is illustrated in Fig. \ref{fig:delay}. As can be observed, to mitigate the near-field beam split effect, the time delay of the TTDs should be designed to compensate for each sub-array the difference in propagation delay to the sub-array experiencing the maximum propagation delay.

\subsection{Finite-Range TTDs} \label{sec:finite}
After obtaining the optimal coefficients for infinite-range TTDs, we now consider practical finite-range TTDs and different configurations.

\subsubsection{Parallel Configuration}
For the parallel configuration, the time delay vector is given by $\mathbf{t} = [t_1,t_2,\dots,t_{Q}]^T$, where the output signal of each TTD exhibits a maximum time delay of $t_{\max}$. In this case, for achieving optimal performance, $t_{\max}$ should be larger than the maximum value of $t_q^{\infty}$, i.e., $(r_{\max}^{\text{sub}} - r_{\min}^{\text{sub}})/c$, where $r_{\min}^{\text{sub}} = \min\{ r_q^{\text{sub}}, \forall q\}$. According to the geometrical relationship illustrated in Fig. \ref{fig:delay} and the triangle inequality, the maximum value of $r_{\max}^{\text{sub}} - r_{\min}^{\text{sub}}$ can be readily obtained by setting $\theta = 0$ or $180^\circ$, and is given by
\begin{equation} \label{condition_parallel_0}
    \max\{r_{\max}^{\text{sub}} - r_{\min}^{\text{sub}}\} = (N - N_{\text{sub}}) d.
\end{equation}
Therefore, the following condition has to hold for achieving the optimal performance for the parallel configuration:
\begin{equation} \label{condition_parallel}
    t_{\max} \ge (N - N_{\text{sub}})d/c.
\end{equation}
Since the beam split effect is caused by non-negligible propagation delay differences across the antenna array aperture, these differences should be compensated by the TTDs as much as possible. Therefore, a heuristic choice of the time delay of each TTD in the parallel configuration is given by
\begin{equation}
    \textcolor{black}{t_q^{\text{parallel}} = \min\{ t_{\max}, t_q^{\infty}\}.} 
\end{equation}

\begin{figure}[t!]
    \centering
    \includegraphics[width=0.4\textwidth]{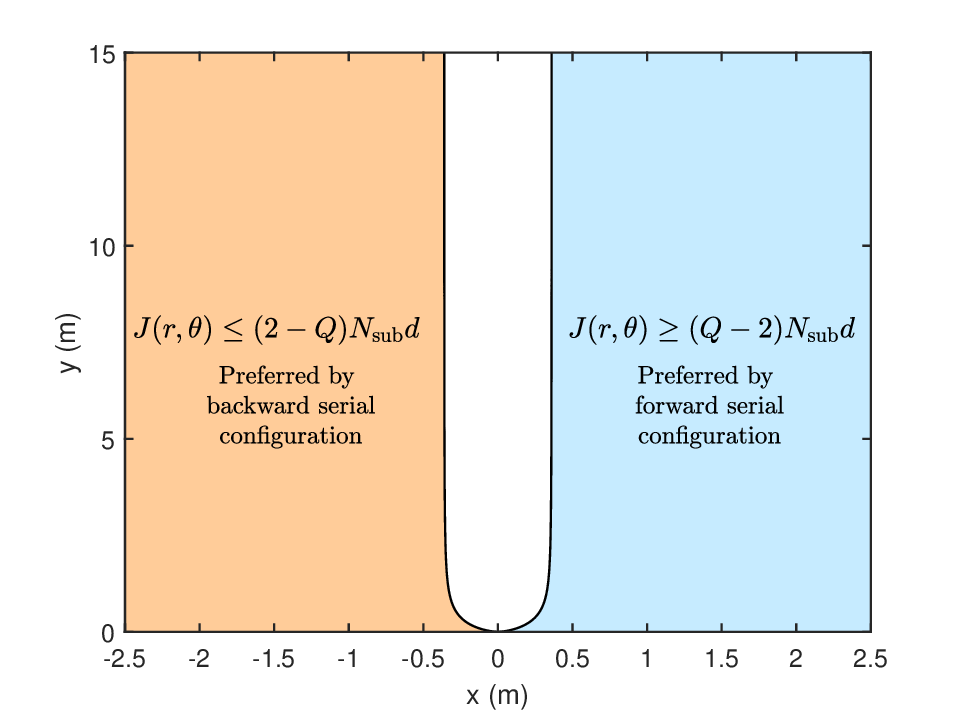}
    \caption{Illustration of the monotonicity regions given in \textbf{Lemma \ref{lemma_1}}. Here, the BS is positioned at $(0,0)$, equipped with $N = 512$ antennas with half-wavelength spacing and operating at $0.1$ THz.}
    \label{fig:region}
\end{figure}

\subsubsection{Serial Configuration} \label{sec:serial}
For the serial configurations, the output time delay of each TTD is subject to monotonicity constraints, as discussed in \textbf{Remark \ref{remark_1}}. Therefore, in the following lemma, we first investigate the monotonicity of the optimal infinite-range time delay $t_q^{\infty}$ with respect to TTD index $q$.
\begin{lemma} \label{lemma_1}
    \emph{Define $J(r, \theta) \triangleq 2 r \cos \theta / \sin^2 \theta$. Time delay $t^{\infty}_q$ \emph{monotonically increases} with respect to $q$ if
    \begin{equation} \label{desired_fs}
        J(r, \theta) \ge  (Q-2) N_{\text{sub}} d,
    \end{equation}
    and it \emph{monotonically decreases} with respect to $q$ if 
    \begin{equation} \label{desired_bs}
        J(r, \theta) \le  (2 - Q) N_{\text{sub}} d.
    \end{equation}
    $t_q^\infty$ \emph{first increases and then decreases} with respect to $q$ if
    \begin{equation}
         (2-Q) N_{\text{sub}} d < J(r, \theta) <  (Q-2) N_{\text{sub}} d,
    \end{equation}
    where the monotonicity changes at the point $q = q_c \triangleq Q/2 + 1 + \lfloor J(r, \theta)/(2N_{\mathrm{\text{sub}}} d)\rfloor$.
    }
\end{lemma}

\begin{IEEEproof}
    Please refer to Appendix \ref{appendiex_1}.
\end{IEEEproof}

\textbf{Lemma \ref{lemma_1}} establishes a connection between the monotonicity of the optimal infinite-range time delay and the location of the user via function $J(r, \theta)$. The monotonicity regions are depicted in Fig. \ref{fig:region}. For the forward serial configuration, the \textcolor{black}{preferred user locations} are defined by $J(r, \theta) \ge (Q-2) N_{\text{sub}} d$, such that the time delay vector maintains the same monotonicity as the optimal infinite-range time delays. Similarly, for the backward serial configuration, the \textcolor{black}{preferred user locations} are defined by $J(r, \theta) \le (2 - Q) N_{\text{sub}} d$.

For the forward serial configuration, the time delay vector can be represented as $\mathbf{t} = [t_1, \sum_{q=1}^2 t_q,\dots,\sum_{q=1}^{Q} t_q ]^T$. \textcolor{black}{When $J(r, \theta) \ge (Q-2) N_{\text{sub}} d$}, the optimal infinite-range time delays follow the order $0 = t_1^{\infty} \le t_2^{\infty} \le \dots \le t_{Q}^{\infty}$. Therefore, the first entry in $\mathbf{t}$ should be set to $t_1=0$. The remaining time delays are determined by the difference of the adjacent optimal infinite-range time delays, and are given by
\begin{align}
    t_q = &\Delta t_q^{\infty} \triangleq t_q^{\infty} - t_{q-1}^{\infty}= \frac{r_{q-1}^{\text{sub}} - r_q^{\text{sub}}}{c}, \forall q = 2,\dots,Q.
\end{align} 
According to the triangle inequality, we have $r_{q-1}^{\text{sub}}  - r_q^{\text{sub}} \le N_{\text{sub}} d$. Consequently, to achieve optimal performance, the maximum achievable time delay of each TTD needs to satisfy the following condition:
\begin{equation} \label{condition_serial}
    t_{\max} \ge N_{\text{sub}} d/c.
\end{equation} 
Compared to condition \eqref{condition_parallel} for the parallel configuration, the above condition is considerably more relaxed since $N_{\text{sub}}$ is much smaller than $N$. When condition \eqref{condition_serial} is not satisfied, the time delay of the TTDs can be designed as $t_1 = 0$ and $t_q = \min\{t_{\max}, \Delta t_q^{\infty} \}, \forall q = 2,\dots,Q$.

However, \textcolor{black}{when $J(r, \theta) < (Q-2) N_{\text{sub}} d$, it becomes impossible to achieve optimal performance due to a mismatch between the monotonicity of the time delays that can be achieved by the forward serial configuration and the monotonicity of $t_q^\infty$.} Consider an example where $\Delta t_q^{\infty} < 0$. In this case, introducing an additional time delay will enlarge the delay difference between the $q$-th and $(q-1)$-th subarrays, thus resulting in a more severe beam split effect. To avoid this issue, the time delay of the $q$-th TTD should be set to $t_q = 0$ if $\Delta t_q^{\infty} \le 0$. In summary, the time delays of the TTDs in the forward serial configuration can be chosen as 
\begin{equation} \label{solution_fs}
    \quad t_q^{\text{f-serial}} = \begin{cases}
        \min\{t_{\max}, \Delta t_q^{\infty}\}, & \Delta t_q^{\infty} \ge 0, \\
        0, & \Delta t_q^{\infty} < 0 \text{ or } q = 1.
    \end{cases}
\end{equation}
Similarly, for the backward serial configuration, the time delay of TTDs can be designed as 
\begin{equation} \label{solution_rs}
    \quad t_q^{\text{b-serial}} = \begin{cases}
        \min\{t_{\max}, -\Delta t_q^{\infty}\}, & \Delta t_q^{\infty} \le 0, \\
        0, & \Delta t_q^{\infty} > 0 \text{ or } q = Q.
    \end{cases}
\end{equation}

\subsubsection{Hybrid Configuration}
For the hybrid configuration, the time delays in the time delay vector first increase and then decrease with index $q$, as shown in \eqref{serial-II}. Therefore, its monotonicity aligns with the monotonicity of the optimal infinite-range time delays for user locations meeting the condition $ (2-Q) N_{\text{sub}} d < J(r, \theta) < (Q-2) N_{\text{sub}} d$. However, the monotonicity change point of the time delays achieved by hybrid configurations is fixed to either $Q/2$ or $Q/2+1$. As a result, the hybrid configuration can achieve optimal performance only if the monotonicity change point of the optimal infinite-range time delays is also located at these two points, i.e., $q_c = Q/2$ or $Q/2+1$, c.f. \textbf{Lemma \ref{lemma_1}}. Given that $q_c =  Q/2 + 1 + \lfloor J(r, \theta)/(2N_{\mathrm{\text{sub}}} d)\rfloor$, it can be shown that the above condition is satisfied if 
\begin{equation} \label{desired_h}
    -2N_{\text{sub}} d \le J(r, \theta) \le 2N_{\text{sub}} d,
\end{equation}
which is the preferred range of the user's location for the hybrid configuration. Furthermore, since the hybrid configuration can be regarded as the combination of two serial configurations, the corresponding time delays can be chosen according to the design of serial configurations in \eqref{solution_fs} and \eqref{solution_rs}, and are given by 
\begin{equation} \label{solution_h}
    t_q^{\text{hybrid}} = \begin{cases}
        t_q^{\text{f-serial}}, & q \le \frac{Q}{2}, \\
        t_q^{\text{b-serial}}, & q > \frac{Q}{2}.
    \end{cases}
\end{equation}

\begin{remark} \label{remark_4}
    \emph{Compared to the serial configuration, \textcolor{black}{the hybrid configuration exhibits a significantly reduced preferred range of user locations}, see \eqref{desired_fs}, \eqref{desired_bs}, and \eqref{desired_h}. However, as evident from the solution in \eqref{solution_h}, half of the TTDs can always operate effectively for maximizing the array gain, i.e., $t_q^{\text{hybrid}} \neq 0$, in the hybrid configuration. In contrast, for the serial configuration, none of the TTDs may be effective if, e.g., the forward serial configuration is chosen but $J(r, \theta) \le (2-Q) N_{\text{sub}} d$.  Therefore, on average the hybrid configuration can achieve a better performance for a single randomly located user, which is confirmed by the numerical results in Section \ref{sec:results}.}
\end{remark}

\section{Multi-User System} \label{sec:multiple}
In this section, we consider a general multi-user system. In contrast to the single-user system, multiple RF chains have to be employed to serve multiple users. In addition, inter-user interference should also be taken into account for the beamforming design for multi-user systems, which makes the problem even more challenging. \textcolor{black}{Given the large number of antenna elements in near-field communications, the complexity of the beamforming design is also critical. Thus, in the following, an efficient low-complexity penalty-based algorithm is developed.}

\subsection{Problem Reformulation}
In multi-user systems, problem \eqref{problem:SE_max} becomes intractable primarily because the analog and digital beamformers are coupled in the complex sum-of-logarithms objective function. To overcome this issue, we introduce the following auxiliary variables:
\begin{equation} \label{equality_Pm}
    \mathbf{P}_m = \mathbf{A} \mathbf{T}_m \mathbf{D}_m, \forall m \in \mathcal{I}(M),
\end{equation}
Let $\mathbf{p}_{m,k}$ denote the vector forming the $k$-th column of $\mathbf{P}_m$.
The achievable communication rate of user $k$ on the $m$-th subcarrier can be rewritten as 
\begin{equation}
    R_{m,k}(\mathbf{P}_m) = \log_2 \left( 1 + \frac{|\mathbf{h}_{m,k}^H \mathbf{p}_{m,k}|^2}{\sum_{i =1, i \neq k}^K |\mathbf{h}_{m,k}^H \mathbf{p}_{m,i}|^2 + \sigma^2 }  \right).
\end{equation}
Then, problem \eqref{problem:SE_max} can be converted into the following equivalent optimization problem:
\begin{subequations} \label{problem:equality}
    \begin{align}
        \max_{\scriptstyle \mathbf{P}_m, \mathbf{A}, \atop \scriptstyle \mathbf{T}_m, \mathbf{D}_m} \quad & \sum_{m=1}^M \sum_{k=1}^K R_{m,k}(\mathbf{P}_m) \\
        \label{cons:new_power}
        \mathrm{s.t.} \quad  & \|\mathbf{P}_m\|_F^2 \le P_t, \forall m \in \mathcal{I}(M), \\
        \label{cons:equality}
        & \mathbf{P}_m = \mathbf{A} \mathbf{T}_m \mathbf{D}_m, \forall m \in \mathcal{I}(M),\\
        \label{cons:analog}
        & \eqref{constraint:analog}-\eqref{constraint:TTD_3}.
    \end{align}
\end{subequations}
In this new problem formulation, the coupling between the different beamformers is shifted to equality constraint \eqref{cons:equality}. However, this new problem is still challenging to solve. To streamline the optimization, we further apply the following two transformations to problem \eqref{problem:equality}, which help to eliminate the constraints.

\textcolor{black}{\subsubsection{Full-power-based Transformation} Following \cite[Proposition 3]{zhao2023rethinking}, it can be proved that any nontrivial stationary point $\mathbf{P}_m$ maximizing $R_{m,k}(\mathbf{P}_m)$ must satisfy the power constraint with equality, i.e., $\|\mathbf{P}_m\|_F^2 = P_t$. Leveraging this property and the fractional form of the objective function, the transmit power constraint in problem \eqref{problem:equality} can be dropped after a simple transformation, leading to the following equivalent optimization problem:
\begin{subequations} \label{problem:equal_power}
    \begin{align}
        \max_{\scriptstyle \mathbf{P}_m, \mathbf{A}, \atop \scriptstyle \mathbf{T}_m, \mathbf{D}_m} \quad & \sum_{m=1}^M \sum_{k=1}^K \tilde{R}_{m,k}(\mathbf{P}_m) \\
        \mathrm{s.t.} \quad &\eqref{constraint:analog}-\eqref{constraint:TTD_3}, \eqref{cons:equality},
    \end{align}
\end{subequations}}
\textcolor{black}{where the new objective function $\tilde{R}_{m,k}(\mathbf{P}_m)$ is given by 
\begin{align}
    &\tilde{R}_{m,k}(\mathbf{P}_m) \nonumber \\ &= \log_2 \left( 1 + \frac{|\mathbf{h}_{m,k}^H \mathbf{p}_{m,k}|^2}{\sum_{i =1, i \neq k}^K |\mathbf{h}_{m,k}^H \mathbf{p}_{m,i}|^2 + \frac{\sigma^2}{P_t} \|\mathbf{P}_m\|_F^2  }  \right).
\end{align}
Let $\mathbf{P}_m^\ddagger$ and $\mathbf{D}_m^\ddagger$ denote stationary solutions of $\mathbf{P}_m$ and $\mathbf{D}_m$ to problem \eqref{problem:equal_power}. Then, according to \cite[Proposition 4]{zhao2023rethinking} and equality constraint \eqref{cons:equality}, the corresponding solutions $\mathbf{P}_m^\star$ and $\mathbf{D}_m^\star$ to the original problem \eqref{problem:equality} can be obtained as follows:
\begin{equation} \label{scale}
    \mathbf{P}_m^\star = \omega \mathbf{P}_m^\ddagger, \quad \mathbf{D}_m^\star = \omega \mathbf{D}_m^\ddagger, 
\end{equation}  
where $\omega = \sqrt{P_t / \|\mathbf{P}_m^\ddagger\|_F^2}$ is a power scaling factor.}

\subsubsection{Penalty-based Transformation}

With this transformation, we aim to address equality constraint \eqref{cons:equality}. First, this constraint can be equivalently reformulated as 
\begin{equation}
    \mathbf{P}_m \mathbf{D}_m^\dagger = \mathbf{A} \mathbf{T}_m, \forall m \in \mathcal{I}(M).
\end{equation}
This simple reformulation helps to streamline the optimization of the analog beamformers $\mathbf{A}$ and $\mathbf{T}_m$, as will be elaborated in the following subsections. Then, the penalty-based method \cite{shi2016joint} is employed, where we incorporate the equality constraint into the objective function as a penalty term, resulting in the following optimization problem:
\begin{subequations} \label{problem:penalty}
    \begin{align}
        \max_{\scriptstyle \mathbf{P}_m, \mathbf{A}, \atop \scriptstyle \mathbf{T}_m, \mathbf{D}_m} \quad & \sum_{m=1}^M \sum_{k=1}^K \tilde{R}_{m,k}(\mathbf{P}_m) - \frac{1}{\rho} \sum_{m=1}^M \left\| \mathbf{P}_m \mathbf{D}_m^\dagger - \mathbf{A} \mathbf{T}_m \right\|_F^2 \\[-0.3em]
        \mathrm{s.t.} \quad & \eqref{constraint:analog}-\eqref{constraint:TTD_3},
    \end{align}
\end{subequations}
where $\rho > 0$ is the penalty factor. The penalty-based method typically employs a double-loop. In the inner loop, problem \eqref{problem:penalty} is solved for a fixed $\rho$.
In the outer loop, penalty factor $\rho$ is initialized with a large value to ensure that the spectral efficiency term is maximized. Then, the value of $\rho$ is gradually decreased in order to effectively drive the penalty term towards zero. This gradual reduction of $\rho$ in the outer loop allows the penalty-based method to strike a balance between optimizing the original objective and satisfying the equality constraints represented by the penalty term. 

Although problem \eqref{problem:penalty} is still non-convex, the optimization variables have been effectively decoupled and the optimization variables are affected by only one constraint. This observation motivates us to exploit the block coordinate descent (BCD) method to solve this problem. In BCD, the optimization variables are divided into several non-overlapped blocks and each block is optimized iteratively while the other blocks are fixed until convergence. For problem \eqref{problem:penalty}, the optimization variables can be divided into three blocks, namely auxiliary variables $\mathbf{P}_m, \forall m$, analog beamforming matrices $\mathbf{A}$ and $\mathbf{T}_m, \forall m$, and digital beamforming matrices $\mathbf{D}_m, \forall m$.     

\textcolor{black}{\subsection{Optimization of Auxiliary Variables}
We start by optimizing auxiliary variables $\mathbf{P}_m, \forall m$, while keeping the other blocks fixed. To address the non-convex sum-of-logarithms objective function, we exploit the weighted minimum mean square error (WMMSE) method \cite{christensen2008weighted}. Specifically, by defining the auxiliary variables $\mathbf{w}_m = [w_{m,1},\dots,w_{m,K}]^T$ and $\mathbf{v}_m = [v_{m,1},\dots,v_{m,K}]^T$, problem \eqref{problem:penalty} can be transformed with respect to $\mathbf{P}_m$ into the following equivalent form:  
\begin{equation} \label{problem:WMMSE}
    \min_{\mathbf{w}_{m}, \mathbf{v}_m, \mathbf{P}_m} \sum_{m=1}^M \sum_{k=1}^K w_{m,k} e_{m,k} + \frac{1}{\rho} \sum_{m=1}^M \left\| \mathbf{P}_m \mathbf{D}_m^\dagger - \mathbf{A} \mathbf{T}_m \right\|_F^2,
\end{equation}
where 
\begin{align}
    e_{m,k} = |v_{m,k}|^2 \left( \sum\nolimits_{i=1}^K |\mathbf{h}_{m,k}^H \mathbf{p}_{m,i}|^2 + \frac{\sigma^2}{P_t} \|\mathbf{P}_m\|_F^2    \right) \nonumber \\ 
    - 2 \mathrm{Re} \left\{ v_{m,k}^* \mathbf{h}_{m,k}^H \mathbf{p}_{m,k}  \right\} + 1.
\end{align}}

\textcolor{black}{For any given $\mathbf{P}_m$, the optimal closed-form solution of $\mathbf{w}_m$ and $\mathbf{v}_m$ to problem \eqref{problem:WMMSE} is given by \cite{christensen2008weighted} 
\begin{align}
    \label{optimal_w}
    w_{m,k} = &2^{ \tilde{R}_{m,k}(\mathbf{P}_m) }, \\
    \label{optimal_v}
    v_{m,k} = &\frac{\mathbf{h}_{m,k}^H \mathbf{p}_{m,k}}{\sum_{i =1}^K |\mathbf{h}_{m,k}^H \mathbf{p}_{m,i}|^2 + \frac{\sigma^2}{P_t} \|\mathbf{P}_m\|_F^2}.
\end{align}}

\textcolor{black}{For any given $\mathbf{w}_{m,k}$ and $\mathbf{v}_{m,k}$, problem \eqref{problem:WMMSE} becomes a convex unconstrained optimization problem with respect to $\mathbf{P}_m$. After some algebraic manipulations, the terms of the objective function of problem \eqref{problem:WMMSE} pertaining to $m$-th subcarrier, which is denoted by $g(\mathbf{P}_m)$, can be reformulated as 
\begin{align}
    g(\mathbf{P}_m) = &\sum_{k=1}^K w_{m,k} e_{m,k} + \frac{1}{\rho} \left\| \mathbf{P}_m \mathbf{D}_m^\dagger - \mathbf{A} \mathbf{T}_m \right\|_F^2 \nonumber \\
    =  &\mathrm{tr}( \mathbf{P}_m^H \mathbf{P}_m \mathbf{\Psi}_m ) + \mathrm{tr}(\mathbf{\Phi}_m \mathbf{P}_m \mathbf{P}_m^H ) \nonumber \\
    & - 2 \mathrm{Re} \{ \mathrm{tr}(\mathbf{\Upsilon}_m \mathbf{P}_m) \} + c_m,
\end{align}
where 
\begin{align}
    \mathbf{\Psi}_m = &\frac{1}{\rho} \mathbf{D}_m^\dagger (\mathbf{D}_m^\dagger)^H, \\
    \mathbf{\Phi}_m = & \sum_{k=1}^K w_{m,k} |v_{m,k}|^2 \left(\mathbf{h}_{m,k} \mathbf{h}_{m,k}^H + \frac{\sigma^2}{P_t} \mathbf{I}_N \right), \\
    \mathbf{\Upsilon}_m = & \sum_{k=1}^K w_{m,k} v_{m,k}^* \mathbf{u}_k \mathbf{h}_{m,k}^H + \frac{1}{\rho} \mathbf{D}_m^\dagger \mathbf{T}_m^H \mathbf{A}^H.
\end{align}
Vector $\mathbf{u}_k$ is a selection vector whose $k$-th entry is one while the other entries are zero, and scalar $c_m$ is a constant scalar unrelated to $\mathbf{P}_m$. The optimal $\mathbf{P}_m$ that minimizes $g(\mathbf{P}_m)$ can be obtained by using the first-order optimality condition. The derivative of $g(\mathbf{P}_m)$ with respect to $\mathbf{P}_m$ is given by 
\begin{equation} \label{Sylvester}
    \frac{\partial g(\mathbf{P}_m)}{\partial \mathbf{P}_m} = \mathbf{\Psi}_m \mathbf{P}_m^H + \mathbf{P}_m^H \mathbf{\Phi}_m - \mathbf{\Upsilon}_m = \mathbf{0}.
\end{equation}      
The above equation is a Sylvester equation, which can be rewritten into the following equivalent form:
\begin{equation}
    \left(\mathbf{\Psi}_m^* \otimes \mathbf{I}_N + \mathbf{I}_K \otimes  \mathbf{\Phi}_m^H \right) \mathrm{vec} \left( \mathbf{P}_m \right) = \mathrm{vec} \left( \mathbf{\Upsilon}_m^H \right).
\end{equation}
Thus, the optimal solution of $\mathbf{P}_m$ is given by 
\begin{equation}
    \mathrm{vec} \left( \mathbf{P}_m \right) = \left(\mathbf{\Psi}_m^* \otimes \mathbf{I}_N + \mathbf{I}_K \otimes  \mathbf{\Phi}_m^H \right)^{-1} \mathrm{vec} \left( \mathbf{\Upsilon}_m^H \right). 
\end{equation} 
To avoid the computationally complex matrix inversion operation in the above solution, the Sylvester equation can also be solved effectively by existing low-complexity methods, such as the Hessenberg-Schur method \cite{golub1979hessenberg}, or with existing toolboxes, such as the MATLAB function $\mathtt{sylvester}$. }

\subsection{Optimization of Analog Beamforming Matrices}
In this subsection, we optimize analog beamforming matrices $\mathbf{A}$ and $\mathbf{T}_m$ for fixed $\mathbf{P}_m$ and $\mathbf{D}_m$. In this case, the optimization problem with respect to $\mathbf{A}$ and $\mathbf{T}_m$ can be simplified as follows
\begin{subequations}
    \begin{align}
        \min_{\mathbf{A}, \mathbf{T}_m} \quad & \sum_{m =1}^M \left\|\mathbf{P}_m \mathbf{D}_m^{\dagger } - \mathbf{A} \mathbf{T}_m \right\|_F^2 \\
        \mathrm{s.t.} \quad & \eqref{constraint:analog}-\eqref{constraint:TTD_3}.
    \end{align}
\end{subequations}
Let $\tilde{\mathbf{p}}_{m,n,q} \in \mathbb{C}^{N_{\text{sub}} \times 1}$ denote a vector comprising entries of matrix $\mathbf{P}_m \mathbf{D}_m^{\dagger}$, specifically the entries from the $((q-1)N_{\text{sub}}+1)$-th to the $q N_{\text{sub}}$-th row of the $n$-th column. Then, the objective function can be further simplified as follows
\begin{align}
    & \left\|\mathbf{P}_m \mathbf{D}_m^{\dagger} - \mathbf{A} \mathbf{T}_m \right\|_F^2 \nonumber \\
    = &\sum_{n=1}^{N_{\text{RF}}} \sum_{q=1}^{Q} \big\| \tilde{\mathbf{p}}_{m,n,q} - \mathbf{a}_{n,q} \mathrm{exp}\left(-j 2\pi f_m \tilde{t}_{n,q}\right) \big\|^2 \nonumber \\
    = &\sum_{n=1}^{N_{\text{RF}}} \sum_{q=1}^{Q} -2 \mathrm{Re} \left\{ \tilde{\mathbf{p}}_{m,n,q}^H  \mathbf{a}_{n,q} \mathrm{exp}\left(-j 2\pi f_m \tilde{t}_{n,q}\right) \right\} + \tilde{c}_m,
\end{align}
where $\tilde{c}_m = \sum_{n=1}^{N_{\text{RF}}} \sum_{q=1}^{Q} ( \tilde{\mathbf{p}}_{m,n,q}^H \tilde{\mathbf{p}}_{m,n,q} + \mathbf{a}_{n,q}^H \mathbf{a}_{n,q} )$.
In particular, $\tilde{c}_m$ is constant due to the fixed value of $\tilde{\mathbf{p}}_{m,n,q}$ as well as the fixed value of $\mathbf{a}_{n,q}^H \mathbf{a}_{n,q} = N_{\text{sub}}$.  

\textcolor{black}{Based on the above results}, for any given $\mathbf{T}_m$, the optimization problem regarding $\mathbf{A}$ can be decomposed into a series of independent optimization problems for each $\mathbf{a}_{n,q}$ as follows:
\begin{subequations}
    \begin{align}
        \max_{\mathbf{a}_{n,q}} \quad & \sum_{m =1}^M \mathrm{Re} \left\{ \tilde{\mathbf{p}}_{m,n,q}^H  \mathbf{a}_{n,q} \mathrm{exp}\left(-j 2\pi f_m \tilde{t}_{n,q}\right) \right\}, \\
        \mathrm{s.t.} \quad & \big|[\mathbf{a}_{n,q}]_i \big| = 1, \forall i \in \mathcal{I}(N_{\text{sub}}).
    \end{align}
\end{subequations}
The optimal solution to the above problem can be readily obtained as follows
\begin{align} \label{solution_A}
    &\mathbf{a}_{n,q} = \mathrm{exp}\left(j \angle \left( \sum\nolimits_{m=1}^M \tilde{\mathbf{p}}_{m,n,q} e^{j 2\pi f_m \tilde{t}_{n,q}} \right) \right).
\end{align}

Similarly, the optimization problem with respect to $\mathbf{T}_m$ for any given $\mathbf{A}$ can also be decomposed into independent subproblems for each time delay vector $\mathbf{t}_n$ as follows
\begin{subequations} \label{problem:TTD}
    \begin{align}
        \max_{\mathbf{t}_n} \quad & \sum_{m =1}^M \sum_{q=1}^{Q} \mathrm{Re} \left\{ \psi_{m,n,q} \mathrm{exp}\left(-j 2\pi f_m \tilde{t}_{n,q}\right) \right\} \\
        \mathrm{s.t.} \quad & \mathbf{t}_n \in \mathcal{T}_n,  \\
        & t_{n, q} \in [0, t_{\max}], \forall q \in \mathcal{I}(Q),
    \end{align}
\end{subequations}
where $\psi_{m,n,q} = \tilde{\mathbf{p}}_{m,n,q}^H  \mathbf{a}_{n,q}$.
In the following, we provide solutions to the above problem for different TTD configurations with different feasible sets $\mathcal{T}_n$. 

\subsubsection{Solution for Parallel Configuration}
For the parallel configuration, we have $\tilde{t}_{n,q} = t_{n,q}$, which implies that the entries of the time delay vector can be optimized individually. Therefore, problem \eqref{problem:TTD} can be converted to 
\begin{subequations}
    \begin{align}
        \max_{t_{n,q}} \quad & \sum_{m =1}^M \mathrm{Re} \left\{ \psi_{m,n,q} \mathrm{exp}\left(-j 2\pi f_m t_{n,q}\right) \right\} \\
        \mathrm{s.t.} \quad & t_{n, q} \in [0, t_{\max}].
    \end{align}
\end{subequations}
Although this problem is non-convex, a near-optimal solution of $t_{n,q}$ can be effectively obtained through a one-dimensional search within the finite interval $[0, t_{\max}]$. Specifically, let $U$ denote the number of search steps, yielding search set $\mathcal{U} =\{0, \frac{t_{\max}}{U-1}, \frac{2t_{\max}}{U-1},\dots,t_{\max}\}$.
Then, a near-optimal solution for $t_{n,q}$ is given by 
\begin{equation} \label{search}
    t_{n,q} = \operatorname*{argmax}_{t_{n,q} \in \mathcal{U}} \sum_{m =1}^M \mathrm{Re} \left\{ \psi_{m,n,q} \mathrm{exp}\left(-j 2\pi f_m t_{n,q}\right) \right\}.
\end{equation} 

\subsubsection{Solution for Serial and Hybrid Configuration}
\textcolor{black}{In contrast to the parallel configuration, for the serial and hybrid configurations, the entries of the time delay vector are coupled and thus cannot be optimized individually. To address the coupling of the delays, we invoke the coordinate descent method. Let us consider the forward serial configuration as an example. When the other entries of $\mathbf{t}_n$ are fixed, the optimization problem for $t_{n,q}$ can be expressed as
\begin{subequations}
    \begin{align}
        \label{obj_serial}
        \max_{t_{n,q}} \quad & \sum_{m =1}^M \sum_{q'=q}^{Q} \mathrm{Re} \left\{ \psi_{m,n,q'} \mathrm{exp}\left(-j 2\pi f_m \sum\nolimits_{i=1}^{q'} t_{n,i}\right) \right\} \\
        \mathrm{s.t.} \quad & t_{n, q} \in [0, t_{\max}].
    \end{align}
\end{subequations}  
Similar to \eqref{search}, a one-dimensional search can be used to find a near-optimal solution for $t_{n,q}$. Then, the entries of the time vector $\mathbf{t}_n$ are successively optimized in a cyclic fashion until the fractional reduction of the objective value falls below a predefined threshold. The optimization problem for the backward serial and hybrid configurations can be solved in the same manner. Thus, we omit the details here.}

\subsection{Optimization of Baseband Digital Beamforming Matrices}
Finally, the digital beamforming matrix $\mathbf{D}_m, \forall m$, is optimized for given values of the other blocks. Similar to the optimization of auxiliary variables $\mathbf{P}_m, \forall m$, the digital beamforming matrix for each subcarrier can be optimized independently, which leads to 
\begin{equation}
    \min_{\mathbf{D}_m} \quad \|\mathbf{P}_m \mathbf{D}_m^{\dagger } - \mathbf{A} \mathbf{T}_m \|_F^2.
\end{equation}
This problem is a well-known least-square problem, which has the following optimal solution:
\begin{equation} \label{solution_Dm}
    \mathbf{D}_m = (\mathbf{A}\mathbf{T}_m)^\dagger \mathbf{P}_m.
\end{equation}

\subsection{Overall Algorithm, Convergence, and Complexity}
To determine whether the equality constraint in problem \eqref{problem:equality} is violated, an indicator of constraint violation is defined as $\xi = \max\{ \|\mathbf{P}_m \mathbf{D}_m^\dagger - \mathbf{A} \mathbf{T}_m \|_{\infty}, \forall m \}$. The algorithm terminates when the value of $\xi$ becomes smaller than a predefined accuracy threshold. The complete algorithm for solving problem \eqref{problem:equality} is summarized in \textbf{Algorithm \ref{alg:penalty}}. For any fixed value of $\rho$, it can be easily proven that the objective value of \eqref{problem:penalty}, achieved by the BCD iterations in the inner loop, is non-decreasing. As a result, based on the findings in \cite{shi2016joint}, it can be concluded that the proposed \textbf{Algorithm \ref{alg:penalty}} converges to a stationary point of problem \eqref{problem:equality}, which is also a stationary point of the original problem \eqref{problem:SE_max}.

Since matrices $\mathbf{A}$, $\mathbf{T}_m$, and $\mathbf{D}_m$ are updated either with a closed-form expression or a low-complexity one-dimensional search, the complexity of \textbf{Algorithm \ref{alg:penalty}} is dominated by updating the auxiliary variables $\mathbf{P}_m$, which requires solving the 
Sylvester equation \eqref{Sylvester}, which has a complexity of $\mathcal{O}(5 N^3/3 + 10 K^3 + 5 N^2 K + 5 N K^2/2)$ if the Hessenberg-Schur method is used \cite{golub1979hessenberg}.

\begin{algorithm}[tb]
    \caption{\textcolor{black}{Penalty-based algorithm for solving \eqref{problem:equality}.}}
    \label{alg:penalty}
    \begin{algorithmic}[1]
        \STATE{initialize penalty factor $\rho \ge 0$ and feasible optimization variables $\mathbf{P}_m, \mathbf{A}, \mathbf{T}_m,$ and $ \mathbf{D}_m, \forall m$}
        \REPEAT
            \REPEAT 
            \STATE{update $\mathbf{w}_m$ and $\mathbf{v}_m, \forall m$ according to \eqref{optimal_w} and \eqref{optimal_v}}
            \STATE{update $\mathbf{P}_m, \forall m$ by the Sylvester equation \eqref{Sylvester}}
            \STATE{update $\mathbf{A}$ according to \eqref{solution_A}}
            \STATE{update $\mathbf{T}_m, \forall m$ via one-dimensional search}
            \STATE{update $\mathbf{D}_m, \forall m$ according to \eqref{solution_Dm}}
            \UNTIL{the fractional reduction of the objective value of \eqref{problem:penalty} falls below a predefined threshold}
        \STATE{update the penalty factor as $\rho = \epsilon \rho$, where $0<\epsilon<1$}
        \UNTIL{the indicator $\xi$ of constraint violation falls below a predefined threshold}
        \STATE{scale $\mathbf{P}_m$ and $\mathbf{D}_m, \forall m$ according to \eqref{scale}}
    \end{algorithmic}
\end{algorithm}

\subsection{Hybrid-Forward-and-Backward Serial Configurations}
In multi-user systems with multiple RF chains, different TTD configurations can be applied for each RF chain, leading to a multitude of possible permutations. The proposed \textbf{Algorithm \ref{alg:penalty}} optimizes the TTD coefficients of each RF chain independently. This approach enables the effective handling of any desired permutation. Nevertheless, the selection of an appropriate permutation plays a crucial role in enhancing overall system performance. In this paper, our objective is to evaluate and compare the performance of the parallel, serial, and hybrid TTD configurations. Thus, we make the assumption that all RF chains are limited to the utilization of one of these configurations. Under this assumption, there is only one possible permutation when employing the parallel and hybrid configurations. On the contrary, in the case of the serial configuration, since it can be implemented either in a forward or backward manner, there exist multiple possible permutations. According to the results in Section \ref{sec:finite}, for multiple RF chains, exploiting only either the forward or the backward implementation may not be suitable for the serial configuration, since it can eliminate the beam split effect across around only half of the entire space of possible user locations based on the results in \textbf{Lemma \ref{lemma_1}}. As a remedy, we propose a hybrid-forward-and-backward (HFB) implementation to extend the coverage of the serial configuration. In the HFB serial configuration, the TTDs for half of the RF chains are configured in the forward serial manner, while the remaining half adopt the backward serial configuration. Thus, we have 
\begin{equation}
    \tilde{t}_{n,q}^{\text{MFB}} = \begin{cases}
        \sum_{i=1}^q t_{n,i}, & n \le \frac{N_{\text{RF}}}{2}, q \in \mathcal{I}(Q), \\
        \sum_{i=q}^Q t_{n,i}, & n > \frac{N_{\text{RF}}}{2}, q \in \mathcal{I}(Q).
    \end{cases}
\end{equation}
Consequently, the power can be adaptively allocated to the two resulting groups of RF chains according to the user locations, thereby enhancing the average performance.

\begin{table}[!t]
    \caption{Simulation parameters.}
    \label{table_para}
    \footnotesize
    \centering
    \begin{tabular}{|c|c|}
    \hline
    Transmit power at the BS $P_t$  & $20$ dBm \\ \hline
    Noise power density & $-174$ dBm/Hz \\ \hline
    Number of antennas at the BS $N$       & $512$ \\ \hline
    System bandwidth $B$  & $10$ GHz \\ \hline 
    Central OFDM frequency $f_c$  & $100$ GHz \\ \hline 
    Number of OFDM subcarriers $M$  & $10$ \\ \hline 
    Length of OFDM cyclic prefix $L_{\text{cp}}$  & $4$ \\ \hline 
    Number of TTDs for each RF chain $Q$ & $32$ \\ \hline 
    Maximum time delay of TTDs $t_{\max}$ & $80$ ps \\ \hline
    Number of channel paths $L_k$ & $4$ \\ \hline  
    Scattering loss $\Lambda_l$ & $-15$ dB \\ \hline
    Transmit and receive antenna gains $G_t, G_r$  & $15$ dB, $5$ dB \\ \hline  
    \end{tabular}
\end{table}

\section{Numerical Results} \label{sec:results}

In this section, numerical results are provided to evaluate the performance of the considered TTD configurations. The users are assumed to be randomly distributed within a range of $5\sim 15$ m from the BS. The number of RF chains is set to $N_{\text{RF}} = 1$ for single-user systems and $N_{\text{RF}} = 4$ for multi-user systems. 
In contrast to low-frequency communications, where pathloss is mainly caused by the spreading loss, in the mmWave and THz bands, the absorption loss is also relevant. Given frequency $f$ and propagation distance $r$, the overall pathloss can be modelled as \cite{jornet2011channel}
\begin{equation}
    \eta_{\text{pathloss}}(f, r) = \left( 4 \pi f r/c \right)^2 \mathrm{exp}\left(k_{\text{abs}}(f) r\right),
\end{equation}
where $k_{\text{abs}}(f)$ denotes the frequency-dependent medium absorption coefficient. The exact value of $k_{\text{abs}}(f)$ can be obtained from the high-resolution transmission (HITRAN) database \cite{rothman2013hitran2012}. Therefore, the channel gain of the LoS component can be modeled as $|\beta_{m,k}|^2 = \eta_{\text{pathloss}}^{-1}(f_m, r_k) G_r G_t$, 
where $G_r$ and $G_t$ denote the antenna gains at the transmitter and receiver, respectively. For the NLoS components, the scattering loss also has to be considered. Therefore, the channel gain of the $l$-th  NLoS component is modeled as $|\tilde{\beta}_{m,k,l}|^2 = \Lambda_{l} \eta_{\text{pathloss}}^{-1}(f_m, \breve{r}_{k,l}) G_r G_t$, 
where $\Lambda_l$ denotes the scattering loss and $\breve{r}_{k,l}$ denotes the overall propagation distance from the BS to user $k$ via the $l$-th scatterer. The simulation parameters provided in Table \ref{table_para} are adopted unless specified otherwise.

For the proposed penalty-based algorithm, the penalty factor is initialized as $\rho=10^4$. The reduction factor the convergence threshold, and the number of search steps for the one-dimensional search are set to $\epsilon = 10^{-1}$, $10^{-4}$, and $U = 10^3$, respectively. The following three benchmark schemes are considered in our simulation:
\begin{itemize}
    \item \textbf{Optimal Full-Digital Beamforming (BF)}: For this benchmark scheme, each antenna is fed by a dedicated RF chain. Therefore, a full-dimensional baseband digital beamformer can be employed for each subcarrier. This benchmark scheme provides a performance upper bound.
    \item \textbf{Optimal TTD-BF}: This benchmark scheme assumes infinite-range TTDs, i.e., $t_{\max} = + \infty$, thus achieving the optimal performance for the TTD-based BF architecture. According to the discussion in \textbf{Remark \ref{remark_2}}, the parallel configuration is adopted for this benchmark scheme.
    \item \textbf{Conventional BF}: This benchmark scheme refers to the PS-only hybrid BF architecture, which can realize only frequency-independent analog beamforming.
\end{itemize}

\begin{figure}[t!]
    \centering
    \includegraphics[width=0.4\textwidth]{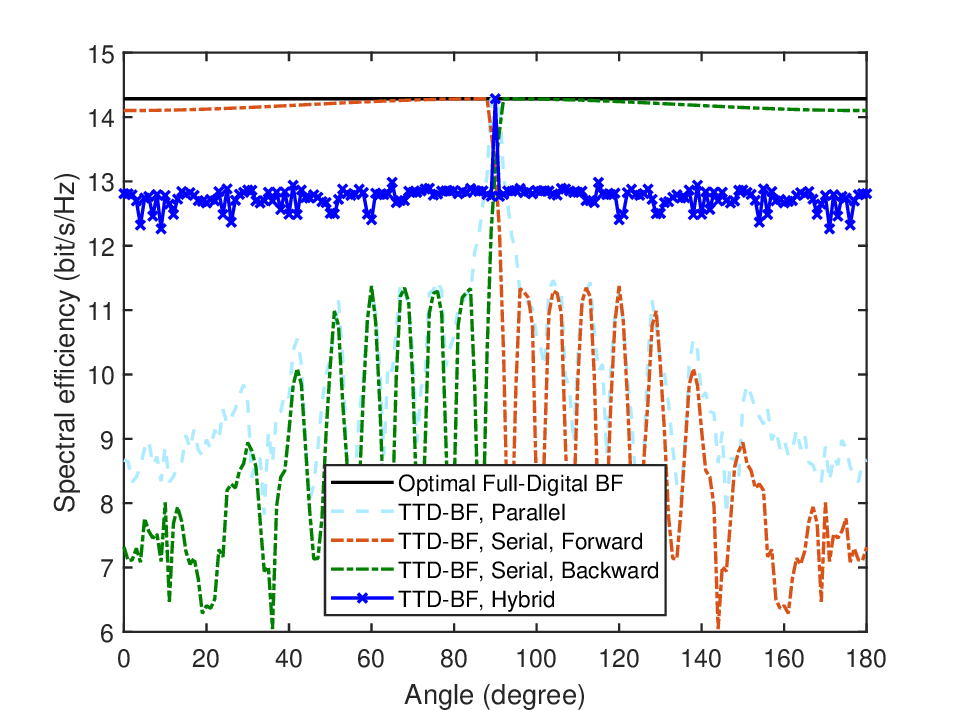}
    \caption{\textcolor{black}{Spectral efficiency versus the direction of a user located at a distance of $10$ m from the BS.}}
    \label{fig_single_angle}
\end{figure}

\begin{figure}[t!]
    \centering
    \includegraphics[width=0.4\textwidth]{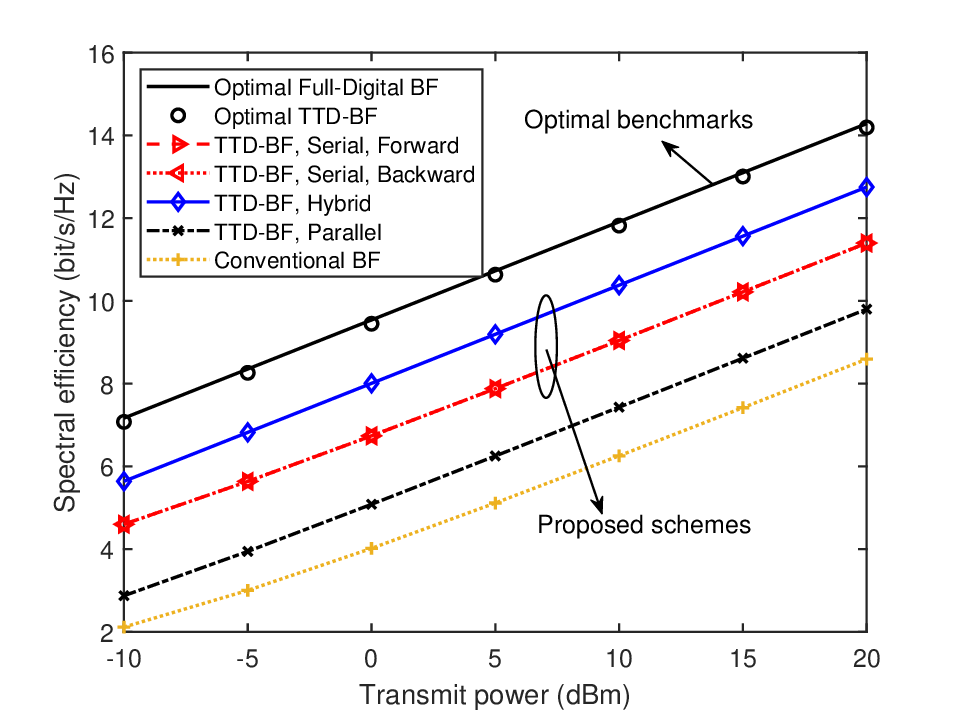}
    \caption{Average spectral efficiency versus the maximum transmit power in the single-user system.}
    \label{fig_single_power}
\end{figure}

\subsection{Single-User System}
We first study the single-user system. To obtain insightful results, only the LoS component of the channel is considered. All results in this subsection are obtained based on the heuristic design in Section \ref{sec:single}. Fig. \ref{fig_single_angle} shows the spectral efficiency achieved by different TTD configurations for a single user located in different directions\footnote{\textcolor{black}{Here, we consider a ULA with optimal scan-angle range for theoretical analysis. In practice, the scan-angle range is typically constrained due to nulls in the element patterns occurring in directions significantly deviating from the broadside \cite{kallnichev2001analysis}. Circular arrays and segmented arrays can be used to address this issue \cite{agiwal2016next, zhang20165g}. Adapting the considered TTD configurations and coefficients to these types of arrays necessitates further research efforts.}}. The distance between the user and the BS is fixed to $10$ m. As can be observed, the parallel configuration attains the optimal performance only when $\theta = 90^\circ$, while resulting in substantial performance loss for other directions. This is because the maximum time delay $t_{\max}$ has to be larger than $2480$ ps for the parallel configuration to achieve optimal performance at the extreme positions of $\theta = 0$ or $180^\circ$, c.f. \eqref{condition_parallel_0} and \eqref{condition_parallel}. For the forward and backward serial configurations, even with a very small maximum time delay, i.e., $t_{\max} = 80$ ps, near-optimal performance can be achieved in one half-space, respectively. However, their performance drastically deteriorates in the other half-space. These results are consistent with our analysis in Section \ref{sec:serial}. Finally, the hybrid configuration yields its peak performance within a narrow range centered around $\theta = 90^\circ$, a range defined by \eqref{desired_h}. Despite this confined optimal region, the hybrid configuration boasts good performance across the entire space as half of the TTDs can always operate effectively. This result is consistent with \textbf{Remark \ref{remark_4}}.

In Fig. \ref{fig_single_power}, we study the average spectral efficiency versus the maximum transmit power for a user located at a fixed distance $10$ m but in a random direction. It can be seen that for infinite-range TTDs, the parallel configuration exhibits comparable performance to full-digital BF. However, when the achievable time delay of the TTDs is limited to $t_{\max} = 80$ ps, the performance of the parallel configuration dramatically decreases to a similar level as conventional BF. In contrast, the performance of the serial and hybrid configurations remains high. Furthermore, for the reasons provided in \textbf{Remark \ref{remark_4}}, the hybrid configuration outperforms the serial configuration in terms of average spectral efficiency.



\subsection{Multi-User System}

\begin{figure}[t!]
    \centering
    \includegraphics[width=0.4\textwidth]{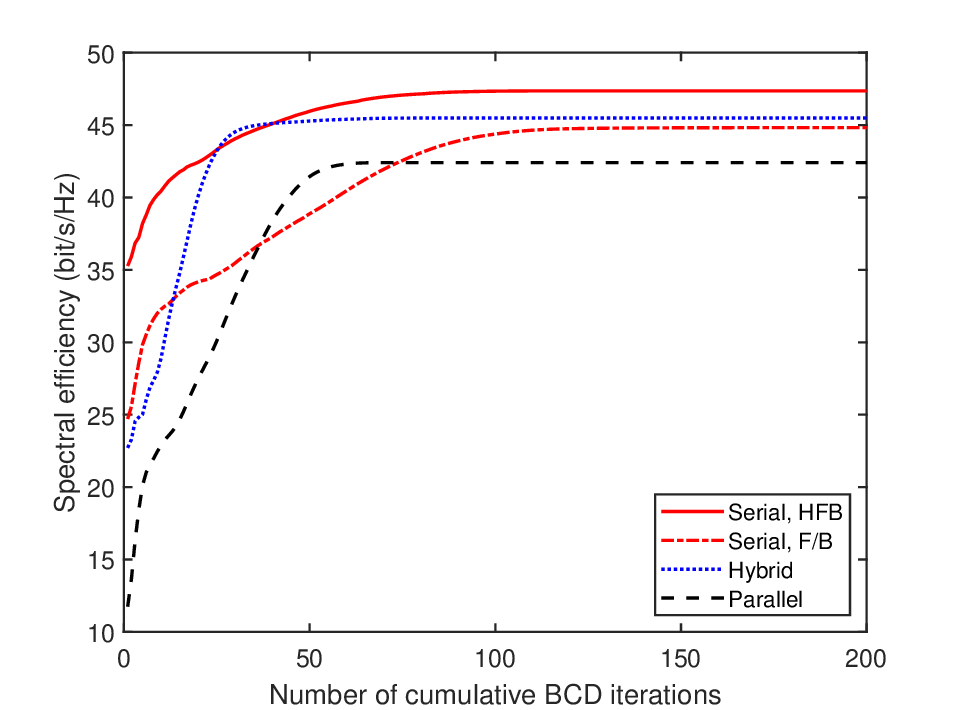}
    \caption{\textcolor{black}{Average convergence behavior of the proposed algorithm in the multi-user system.}}
    \label{fig_convergence}
    \vspace{-0.35cm}
\end{figure}

\begin{figure}[t!]
    \centering
    \includegraphics[width=0.4\textwidth]{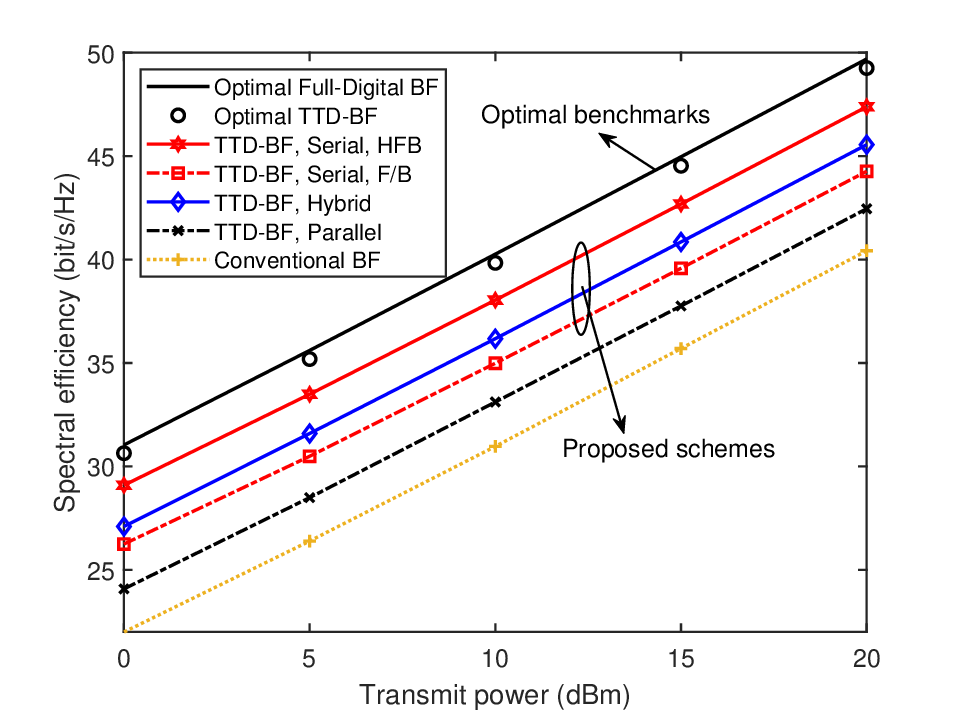}
    \caption{Average spectral efficiency versus maximum transmit power in the multi-user system.}
    \label{fig_rate_power}
\end{figure}

Next, we study a multi-user system with $K = 4$ users and $N_{\text{RF}} = 4$ RF chains. All following results are obtained by averaging over $100$ channel realizations of the four users, which are located randomly within a distance range of $5$ to $15$ m from the BS. 
\textcolor{black}{Fig. \ref{fig_convergence} demonstrates the convergence behavior of the proposed algorithm. Here, “Serial, F/B” implies that the TTDs for all RF chains are configured either in a forward or backward manner\footnote{\textcolor{black}{When utilizing either the forward or backward serial configurations for all RF chains in multi-user systems, the average performance for randomly positioned users is the same because the forward and backward configurations are mirror images of each other with respect to the normal line of the antenna array. Therefore, for our simulation results, we regarded them as the same configuration.} }. As can be observed, the spectral efficiency gradually increases as the iterations proceed and finally converges to a stable value for all considered TTD configurations, which confirms the effectiveness of the proposed algorithm.}

Fig. \ref{fig_rate_power} illustrates the relationship between average spectral efficiency and maximum transmit power in multi-user systems. It is interesting to observe that while the hybrid configuration demonstrates superior performance compared to the F/B serial configuration, it fails to surpass the performance of the HFB serial configuration. This result is attributed to two key factors. Firstly, the time delay achieved by the serial configuration is much larger than the hybrid configuration. Secondly, the hybrid utilization of forward and backward serial configurations substantially extends the effective coverage area. Finally, the performance gain of the parallel configuration over conventional hybrid beamforming is marginal, due to the limited time delay it achieves.

\begin{figure}[t!]
    \centering
    \includegraphics[width=0.4\textwidth]{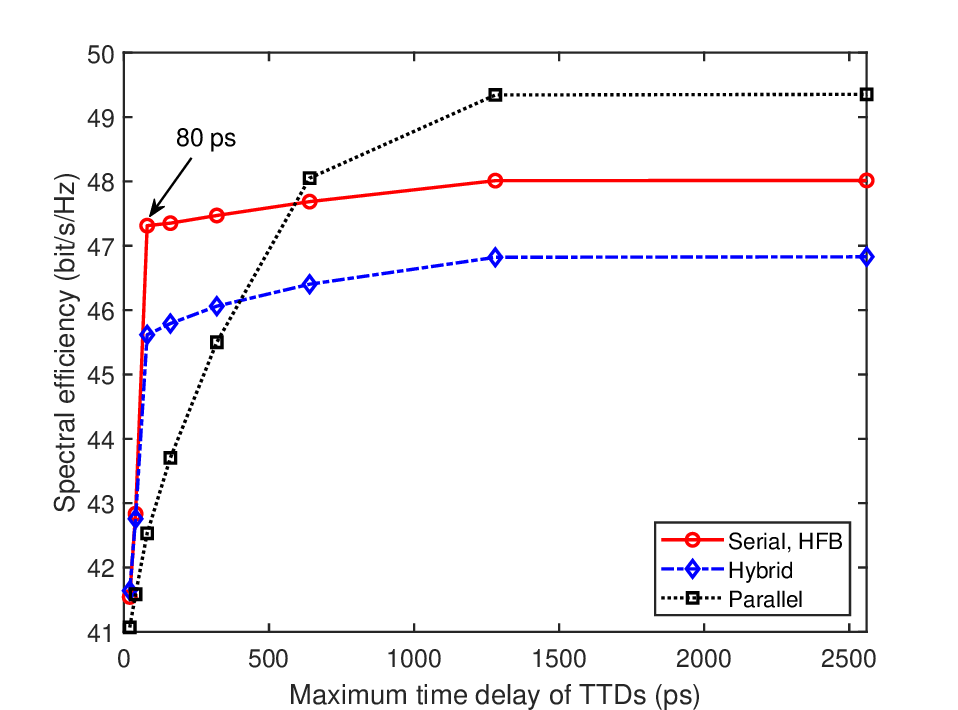}
    \caption{Average spectral efficiency versus maximum time delay $t_{\max}$ of the TTDs in the multi-user system.}
    \label{fig_rate_tmax}
    \vspace{-0.4cm}
\end{figure}

\begin{figure}[t!]
    \centering
    \includegraphics[width=0.4\textwidth]{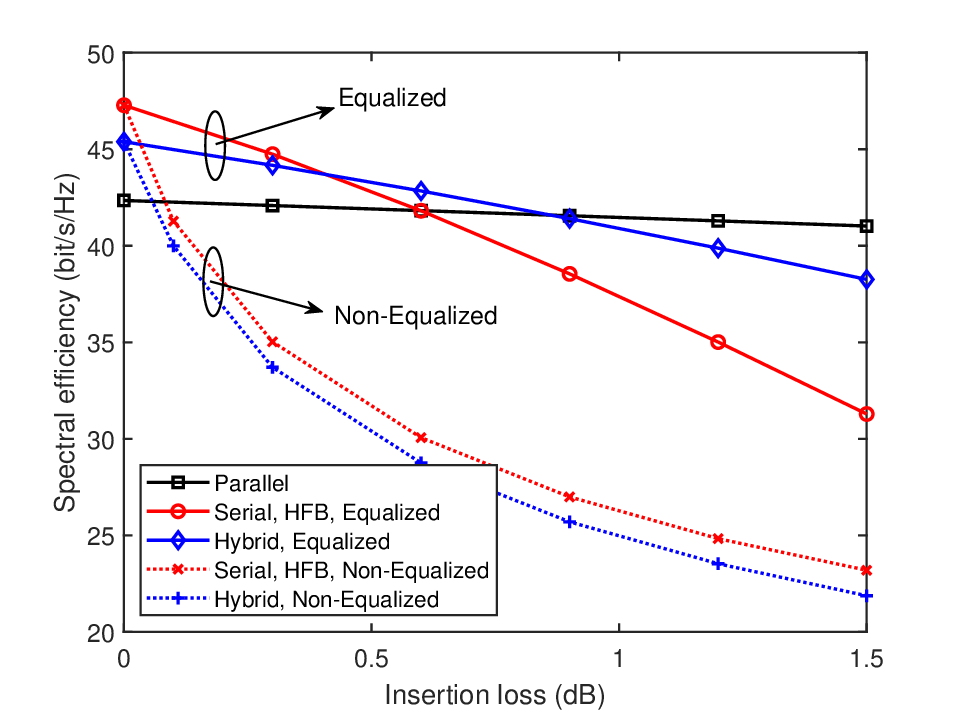}
    \caption{Average spectral efficiency versus insertion loss $\eta$ in the multi-user system.}
    \label{fig_loss}
\end{figure}

\begin{table*}[t!]
    \caption{\textcolor{black}{Summary of different TTD configurations}}
    \label{table_compare}
    \resizebox{\linewidth}{!}{
    \centering
    \begin{tabular}{|c|c|c|c|c|}
        \hline
        \textbf{Configuration}  &\textbf{Preferred Location of Users}   & \textbf{Required Delay of TTDs} & \textbf{Insertion Loss} &\textbf{Suitable Scenario}
        \\ \hline
        Serial & \makecell[l]{Forward: $J(r_k, \theta_k) \ge (Q-2)N_{\text{sub}} d$ \\Backward: $J(r_k, \theta_k) \le (2-Q)N_{\text{sub}} d$} &Small: $N_{\text{sub}}d/c$&High: $\frac{1 - \eta^Q}{(1 - \eta) Q} \times \eta$& \makecell[l]{Short-range TTDs, \\ multi-user systems} \\ \hline 
        Hybrid &$-2N_{\text{sub}} d \le J(r_k, \theta_k) \le 2N_{\text{sub}} d$  &Small: $N_{\text{sub}}d/c$ &Medium: $\frac{1 - \eta^{\frac{Q}{2}}}{(1 - \eta) \frac{Q}{2}} \times \eta$ & \makecell[l]{Short-range TTDs, \\ single-user systems} \\ \hline
        Parallel & -  &Large: $(N-N_{\text{sub}}) d/c$&Low: $\eta$ & \makecell[l]{Long-range TTDs} \\ \hline 
    \end{tabular}
    }
\end{table*}

Fig. \ref{fig_rate_tmax} shows the average spectral efficiency versus the maximum time delay $t_{\max}$ of the TTDs. All three configurations demonstrate enhanced performance as $t_{\max}$ increases. The parallel configuration necessitates a minimum achievable time delay of $500$ ps to attain a performance comparable to that of the serial configuration, affirming the advantages of the serial setup for finite-range TTDs. However, as the maximum time delay becomes sufficiently large, the parallel configuration emerges as the superior choice due to the absence of monotonicity constraints. Furthermore, when the maximum time delay falls below $N_{\text{sub}} d/c = 80$ ps, a notable performance decline is observed in the serial configuration. This result implies that satisfying the condition stated in \eqref{condition_serial} is essential to ensure optimal performance for the serial configuration.

Fig. \ref{fig_loss} depicts the impact of the insertion loss $\eta$ of the TTDs and power splitters. \textcolor{black}{The coefficients of the power splitters for the non-equalized and equalized schemes are designed based on \eqref{solution_non_equ} and \eqref{solution_eqn}, respectively.} The importance of equalization is evident for the serial and hybrid configurations, as it ensures the effective utilization of each individual TTD. Moreover, compared to the parallel configuration, the serial and hybrid configurations are more sensitive to insertion loss due to the accumulation of the losses of several TTDs and power splitters. \textcolor{black}{In particular, the serial and hybrid configurations yield inferior performance compared to the parallel configuration for $\eta > 0.6$ dB and $\eta > 0.9$ dB, respectively. This underscores the critical role of low-loss TTDs and power splitters for the effectiveness of the serial and hybrid configurations. Although there are low-loss TTDs and power splitters with insertion losses of $0.07$ dB \cite{xiang2017low} and $0.15$ dB \cite{du2016unequal}, respectively, they are only applicable to low-frequency bands. The development of low-loss TTDs and power splitters for high-frequency bands remains an open problem, necessitating further research.}

\vspace{-0.2cm}
\section{Conclusion} \label{sec:conclusion}
In this paper, we introduced and examined two different TTD configurations, namely the serial and hybrid configurations, as building blocks for hybrid beamforming architectures to maximize the performance of near-field communication systems. The principal objective behind these TTD configurations was to curtail the demands on the hardware associated with TTDs. In conjunction with this, we addressed the challenge of mitigating the cumulative insertion losses caused by the serial and hybrid configurations by devising a corresponding equalization strategy. A comprehensive investigation of beamforming design was provided, encompassing both single-user and multi-user scenarios. For the single-user case, we analytically derived a closed-form solution. In contrast, the multi-user scenario demanded a more intricate design, for which we proposed a penalty-based iterative algorithm. This algorithm facilitated the attainment of stationary-point solutions. Moreover, we enhanced the performance of the serial TTD configuration through an HFB approach. Numerical results were presented to validate the efficacy of the proposed configurations. Based on the results obtained in this paper, the main characteristics of different TTD configurations are summarized in Table \ref{table_compare}, which may serve as a guide for future works. \textcolor{black}{Finally, multi-carrier OFDM modulation was taken into account to address the frequency-selective effect in this paper. The inherent high peak-to-average power ratio (PAPR) associated with OFDM modulation can significantly degrade transmitter efficiency, especially for high frequencies. Hence, studying the performance of the proposed TTD configurations for low-PAPR single-carrier wideband systems \cite{4607215, 9790794} constitutes a promising research direction.}
\begin{appendices}
    \section{Proof of Lemma \ref{lemma_1}} \label{appendiex_1}

    To simplify the analysis, we exploit the Fresnel approximation to approximate distance $r_q^{\text{sub}}$. This leads to \cite{liu2023near_tutorial}
    \begin{equation}
        r_q^{\text{sub}} \approx r - \chi_q d \cos \theta + \frac{\chi_q^2 d^2 \sin^2 \theta}{2r}.
    \end{equation} 
    Then, the difference between adjacent $t^{\infty}_q, \forall q = 2,\dots,Q$, can be approximated as follows 
    \begin{align}
        \Delta t^{\infty}_q = &\frac{r_{q-1}^{\text{sub}} - r_q^{\text{sub}}}{c} \nonumber \\
        \approx & \frac{N_{\text{sub}}d \cos \theta}{c} - \frac{(2q-2-Q)N_{\text{sub}}^2  d^2 \sin^2 \theta}{2rc}.
    \end{align}
    Sequence $t^{\infty}_q$ is monotonically increasing if $\Delta t^{\infty}_q \ge 0, \forall q = 2,\dots,Q$, which yields
    \begin{align}
        \frac{N_{\text{sub}}d \cos \theta}{c} \ge & \max_q \left\{\frac{(2q-2-Q)N_{\text{sub}}^2  d^2 \sin^2 \theta}{2rc} \right\} \\
        \Leftrightarrow \quad J(r, \theta)  \ge &(Q-2) N_{\text{sub}} d.
    \end{align}
    Similarly, we have $\Delta t_q^{\infty} \le 0, \forall q = 2,\dots,Q$ if $J(r, \theta) \le (2-Q) N_{\text{sub}} d$, which implies that $t_q^{\infty}$ is monotonically decreasing in this region. It is also easy to prove that if $(2-Q) N_{\text{sub}} d < J(r, \theta) < (Q-2) N_{\text{sub}} d$, $t_q^{\infty}$ first increases and then decreases with respect to $q$. In this region, the monotonicity of $t_q^{\infty}$ changes when $\Delta t_q^{\infty} = 0$, yielding $q = Q/2 + 1 + J(r, \theta)/(2N_{\mathrm{\text{sub}}} d)$. Since $q$ has to an integer, the monotonicity change point $q_c$ is given by $q_c =  Q/2 + 1 + \lfloor J(r, \theta)/(2N_{\mathrm{\text{sub}}} d)\rfloor$. The proof is thus complete.

\end{appendices}

\bibliographystyle{IEEEtran}
\bibliography{reference/mybib}

\end{document}